\documentclass[preprint]{ptephy}

\preprintnumber{KYUSHU-HET-140}

\usepackage{amssymb}
\usepackage{amsthm}
\usepackage{amsmath}
\usepackage{booktabs}
\DeclareMathOperator{\tr}{tr}

\DeclareMathOperator{\re}{Re}

\newcommand{\Slash}[1]{{\ooalign{\hfil/\hfil\crcr$#1$}}}
\numberwithin{equation}{section}

\usepackage{url}

\begin{document}

\title{Lattice energy--momentum tensor from the Yang--Mills gradient flow---%
inclusion of fermion fields}

\author{\name{\fname{Hiroki} \surname{Makino}}{1},
\name{\fname{Hiroshi} \surname{Suzuki}}{1,\ast}
}

\address{%
\affil{1}{Department of Physics, Kyushu University, 6-10-1 Hakozaki, Higashi-ku, Fukuoka, 812-8581, Japan}
\email{hsuzuki@phys.kyushu-u.ac.jp}
}

\begin{abstract}
Local products of fields deformed by the so-called Yang--Mills gradient flow
become renormalized composite operators. This fact has been utilized to
construct a correctly normalized conserved energy--momentum tensor in the
lattice formulation of the pure Yang--Mills theory. In the present paper, this
construction is further generalized for vector-like gauge theories containing
fermions.
\end{abstract}
\subjectindex{B01, B31, B32, B38}
\maketitle

\section{Introduction and the main result}
\label{sec:1}
Energy and momentum are fundamental notions in physics. In lattice field theory
(the best-developed non-perturbative formulation of quantum field theory),
however, the construction of the corresponding Noether current, the
energy--momentum
tensor~\cite{Callan:1970ze,Coleman:1970je,Freedman:1974gs,Joglekar:1975jm}, is
not straightforward~\cite{Caracciolo:1988hc,Caracciolo:1989pt}, because the
translational invariance is explicitly broken by the lattice structure. In a
recent paper~\cite{Suzuki:2013gza}, a possible method to avoid this
complication being inherent in the energy--momentum tensor on the lattice has
been proposed on the basis of the Yang--Mills gradient flow (or the Wilson flow
in the context of lattice gauge
theory)~\cite{Luscher:2010iy,Luscher:2011bx,Luscher:2013cpa}.%
\footnote{In~Refs.~\cite{Giusti:2010bb,Giusti:2012yj,Robaina:2013zmb,%
Giusti:2013sqa,Giusti:2013mxa,Giusti:2014ila}, an interesting method to define
a lattice energy--momentum tensor from shifted boundary conditions has been
developed. In~Ref.~\cite{Suzuki:2012wx}, a method on the basis of the
$\mathcal{N}=1$ supersymmetry has been proposed.} See
Ref.~\cite{Luscher:2013vga} for a recent review on the gradient flow and
Refs.~\cite{Borsanyi:2012zs,Borsanyi:2012zr,Fodor:2012td,Fritzsch:2013je,%
Monahan:2013lwa,Shindler:2013bia,Bonati:2014tqa} for its applications in
lattice gauge theory.

The basic idea of~Ref.~\cite{Suzuki:2013gza} is the following:\footnote{The
following reasoning was inspired by pioneering experimentation by E.~Itou
and~M.~Kitazawa (unpublished).} Consider the pure Yang--Mills theory. The
gradient flow deforms the bare gauge field~$A_\mu(x)$ according to a flow
equation with a flow time~$t$ [Eq.~\eqref{eq:(3.1)} below] and this makes gauge
field configurations ``smooth'' for positive flow times. It can then be shown
to all orders in perturbation theory that any local products of the flowed
gauge field~$B_\mu(t,x)$ for any strictly positive flow time~$t$ is ultraviolet
(UV) finite when expressed in terms of renormalized
parameters~\cite{Luscher:2011bx}. In particular, no multiplicative
renormalization factor is required to make those local products finite. In
other words, they are renormalized composite operators. Such UV-finite
quantities should be ``universal'' in the sense that they are independent of
the UV regularization chosen, in the limit in which the regulator is removed.
This suggests a possibility that by using the gradient flow as an intermediate
tool one may bridge composite operators defined with the dimensional
regularization, with which the translational invariance is
manifest,\footnote{The drawback of the dimensional regularization is, of
course, that it is defined only in perturbation theory.} and those in the
lattice regularization with which one may carry out non-perturbative
calculations.\footnote{We thus implicitly assume that the finiteness of the
flowed fields that can rigorously be proven only in perturbation theory
persists even in the non-perturbative level.} Following this idea, a formula
that expresses the correctly normalized conserved energy--momentum tensor as a
$t\to0$ limit of a certain combination of the flowed gauge field was
derived~\cite{Suzuki:2013gza}. This formula provides a possible method to
compute correlation functions of the energy--momentum tensor by using the
lattice regularization because the universal combination in the formula should
be independent of the regularization. An interesting point is that the small
flow-time behavior of the (universal) coefficients in the formula can be
determined by perturbation theory thanks to the asymptotic freedom. This
implies that if the lattice spacing is fine enough a further non-perturbative
determination of the coefficients is not necessary. (Practically,
non-perturbative determination of those coefficients may be quite useful and
how this determination can be carried out has been investigated
in~Ref.~\cite{DelDebbio:2013zaa}.) Although the validity of the formula
in~Ref.~\cite{Suzuki:2013gza}, especially the restoration of the conservation
law in the continuum limit, still remains to be carefully investigated, the
measurement of the interaction measure (the trace anomaly) and the entropy
density of the Yang--Mills theory at finite temperature on the basis of the
formula~\cite{Asakawa:2013laa} shows encouraging results; the method appears to
be promising even practically.

In~Ref.~\cite{Suzuki:2013gza}, the method was developed only for the pure
Yang--Mills theory. It is then natural to ask for wider application if the
method can be generalized to gauge theories containing matter fields,
especially fermion fields. In the present paper, we work out this
generalization. We thus suppose a vector-like gauge theory\footnote{We assume
that the theory is asymptotically free.} with a gauge group~$G$ that contains
\begin{equation}
   \text{$N_{\mathrm{f}}$ Dirac fermions in the gauge representation~$R$}.
\label{eq:(1.1)}
\end{equation}
For simplicity, we assume that all $N_{\mathrm{f}}$ fermions possess a common mass; this
restriction might be appropriately relaxed.

Our main result is Eq.~\eqref{eq:(4.70)}, and a step-by-step derivation of this
master formula is given in subsequent sections. For those who are interested
mainly in the final result, here we give a brief explanation of how to read the
master formula~\eqref{eq:(4.70)}: The left-hand side is the
correctly normalized conserved energy--momentum tensor (with the vacuum
expectation value subtracted); our formula~\eqref{eq:(4.70)} holds only when
the energy--momentum tensor is separated from other operators in correlation
functions in the position space. The combinations in the right-hand side are
defined by~Eqs.~\eqref{eq:(4.1)}--\eqref{eq:(4.5)}. There, $G_{\mu\nu}^a(t,x)$
and~$D_\mu$ are the field strength and the covariant derivative of the flowed
gauge field, respectively (the definition of the flowed gauge field is
identical to that
of~Refs.~\cite{Luscher:2010iy,Luscher:2011bx,Luscher:2013cpa}); our
\emph{ringed\/} flowed fermion fields
$\mathring{{\chi}}(t,x)$ and~$\mathring{\Bar{\chi}}(t,x)$ are, on the other
hand, somewhat different from those of~Ref.~\cite{Luscher:2013cpa}, $\chi(t,x)$
and~$\Bar{\chi}(t,x)$, and they are related by~Eqs.~\eqref{eq:(3.20)}
and~\eqref{eq:(3.21)}. The coefficient functions $c_i(t)$
in~Eq.~\eqref{eq:(4.70)} are given
by~Eqs.~\eqref{eq:(4.72)}--\eqref{eq:(4.76)}. There, $\Bar{g}(q)$ and
$\Bar{m}(q)$ are the running coupling and the running mass parameter defined by
Eqs.~\eqref{eq:(4.20)} and~\eqref{eq:(4.21)}, respectively; throughout this
paper, $m$ denotes the (common) renormalized mass of the fermions.
Equations~\eqref{eq:(4.72)}--\eqref{eq:(4.76)} are for the minimal subtraction
($\text{MS}$) scheme and the result for the modified minimal subtraction
($\overline{\text{MS}}$) scheme can be obtained by the
replacement~\eqref{eq:(4.77)}. $b_0$ and~$d_0$ are the first coefficients of
renormalization group functions, Eq.~\eqref{eq:(2.16)}
and~Eq.~\eqref{eq:(2.18)}, respectively. The energy--momentum tensor is given
by the $t\to0$ limit in the right-hand side of~Eq.~\eqref{eq:(4.70)}. As in the
pure Yang--Mills case mentioned above, the combination in the right-hand side
is UV finite and one may use the lattice regularization to compute the
correlation functions of the combination in the right-hand side. In this way,
correlation functions of the correctly normalized conserved energy--momentum
tensor are obtained. To ensure the ``universality,'' however, the continuum
limit has to be taken before the $t\to0$ limit. Practically, with a finite
lattice spacing~$a$, the flow time~$t$ cannot be taken arbitrarily small to
keep the contact with the continuum physics. Instead, we have a natural
constraint,
\begin{equation}
   a\ll\sqrt{8t}\ll R,
\label{eq:(1.2)}
\end{equation}
where~$R$ denotes a typical physical scale, such as the hadronic scale or the
box size. The extrapolation for~$t\to0$ thus generally requires a sufficiently
fine lattice.

Here is our definition of the quadratic Casimir operators: We set the
normalization of anti-Hermitian generators~$T^a$ of the representation~$R$
as~$\tr_R(T^aT^b)=-T(R)\delta^{ab}$ and~$T^aT^a=-C_2(R)1$. We also denote
$\tr_R(1)=\dim(R)$. From the structure constants in~$[T^a,T^b]=f^{abc}T^c$, we
define $f^{acd}f^{bcd}=C_2(G)\delta^{ab}$. For example, for the fundamental
$N$~representation of~$SU(N)$ for which $\dim(N)=N$, the conventional
normalization is
\begin{equation}
   C_2(SU(N))=N,\qquad T(N)=\frac{1}{2},\qquad
   C_2(N)=\frac{N^2-1}{2N}.
\label{eq:(1.3)}
\end{equation}

\section{Energy--momentum tensor with dimensional regularization}
\label{sec:2}
The description of the energy--momentum tensor in gauge
theory~\cite{Freedman:1974gs,Joglekar:1975jm} is particularly simple with the
dimensional regularization.\footnote{Ref.~\cite{Collins:1984xc} is a very nice
exposition of the dimensional regularization.} This is because this
regularization manifestly preserves the (vectorial) gauge symmetry and the
translational invariance. Thus, in this section, we briefly recapitulate basic
facts concerning the energy--momentum tensor on the basis of the dimensional
regularization.

The action of the system under consideration in a $D$~dimensional Euclidean
space is given by
\begin{equation}
   S=\frac{1}{4g_0^2}\int\mathrm{d}^Dx\,F_{\mu\nu}^a(x)F_{\mu\nu}^a(x)
   +\int\mathrm{d}^Dx\,\Bar{\psi}(x)(\Slash{D}+m_0)\psi(x),
\label{eq:(2.1)}
\end{equation}
where $g_0$ and~$m_0$ are bare gauge coupling and the mass parameter,
respectively. The field strength is defined by
\begin{equation}
   F_{\mu\nu}(x)
   =\partial_\mu A_\nu(x)-\partial_\nu A_\mu(x)+[A_\mu(x),A_\nu(x)],
\label{eq:(2.2)}
\end{equation}
for $A_\mu(x)=A_\mu^a(x)T^a$ and~$F_{\mu\nu}(x)=F_{\mu\nu}^a(x)T^a$, and the
covariant derivative on the fermion is
\begin{equation}
   D_\mu=\partial_\mu+A_\mu.
\label{eq:(2.3)}
\end{equation}
Here, and in what follows, the summation over $N_{\mathrm{f}}$ fermion flavors is always
suppressed. Our gamma matrices are Hermitian and for the trace over the spinor
index we set $\tr 1=4$ for any~$D$. We also set
\begin{equation}
   D=4-2\epsilon.
\label{eq:(2.4)}
\end{equation}

Assuming the dimensional regularization, one can derive a Ward--Takahashi
relation associated with the translational invariance straightforwardly. We
consider the following infinitesimal variation of integration variables in the
functional integral:
\begin{equation}
   \delta A_\mu(x)=\xi_\nu(x)F_{\nu\mu}(x),\qquad
   \delta\psi(x)=\xi_\mu(x)D_\mu\psi(x).
\label{eq:(2.5)}
\end{equation}
Then, since the action changes as\footnote{Here, to make the perturbation
theory well defined, we implicitly assume the existence of the gauge-fixing
term and the Faddeev--Popov ghost fields. However, since they do not explicitly
appear in correlation functions of gauge-invariant operators, we neglect these
elements in what follows.}
\begin{equation}
   \delta S=-\int\mathrm{d}^Dx\,\xi_\nu(x)\partial_\mu T_{\mu\nu}(x),
\label{eq:(2.6)}
\end{equation}
where the energy--momentum tensor~$T_{\mu\nu}(x)$ is defined by
\begin{align}
   T_{\mu\nu}(x)
   &\equiv\frac{1}{g_0^2}\left[
   F_{\mu\rho}^a(x)F_{\nu\rho}^a(x)
   -\frac{1}{4}\delta_{\mu\nu}F_{\rho\sigma}^a(x)F_{\rho\sigma}^a(x)
   \right]
\notag\\
   &\qquad{}
   +\frac{1}{4}
   \Bar{\psi}(x)\left(\gamma_\mu\overleftrightarrow{D}_\nu
   +\gamma_\nu\overleftrightarrow{D}_\mu\right)\psi(x)
   -\delta_{\mu\nu}\Bar{\psi}(x)
   \left(\frac{1}{2}\overleftrightarrow{\Slash{D}}
   +m_0\right)\psi(x),
\label{eq:(2.7)}
\end{align}
with
\begin{equation}
   \overleftrightarrow{D}_\mu\equiv D_\mu-\overleftarrow{D}_\mu,\qquad
   \overleftarrow{D}_\mu\equiv\overleftarrow{\partial}_\mu-A_\mu,
\label{eq:(2.8)}
\end{equation}
we have
\begin{equation}
   \left\langle\mathcal{O}_{\text{out}}
   \int_{\mathcal{D}}
   \mathrm{d}^Dx\,\partial_\mu T_{\mu\nu}(x)\,\mathcal{O}_{\text{in}}\right\rangle
   =-\left\langle\mathcal{O}_{\text{out}}\,\partial_\nu\mathcal{O}_{\text{in}}
   \right\rangle,
\label{eq:(2.9)}
\end{equation}
where $\mathcal{O}_{\text{in}}$ ($\mathcal{O}_{\text{out}}$) is a collection of
gauge-invariant operators localized inside (outside) the finite integration
region~$\mathcal{D}$. This relation shows that the energy--momentum tensor
generates the infinitesimal translation and, at the same time, the bare
quantity~\eqref{eq:(2.7)} does not receive the multiplicative renormalization.
Thus, we define a renormalized finite energy--momentum tensor by subtracting
its (possibly divergent) vacuum expectation value:
\begin{equation}
   \left\{T_{\mu\nu}\right\}_R(x)
   \equiv T_{\mu\nu}(x)-\left\langle T_{\mu\nu}(x)\right\rangle.
\label{eq:(2.10)}
\end{equation}

A fundamental property of the energy--momentum tensor is the trace
anomaly~\cite{Crewther:1972kn,Chanowitz:1972vd,Adler:1976zt,Nielsen:1977sy,%
Collins:1976yq}. One simple way to derive
this~\cite{Fujikawa:1980rc,Fujikawa:2004cx} is to set $\xi_\mu(x)\propto x_\mu$
in~Eq.~\eqref{eq:(2.5)} and compare the resulting relation with the
renormalization group equation. After some consideration, this yields
\begin{equation}
   \delta_{\mu\nu}
   \left\{T_{\mu\nu}\right\}_R(x)
   =-\frac{\beta}{2g^3}\left\{F_{\rho\sigma}^aF_{\rho\sigma}^a\right\}_R(x)
   -(1+\gamma_m)m\left\{\Bar\psi\psi\right\}_R(x),
\label{eq:(2.11)}
\end{equation}
where we assume that the renormalized operators in the right-hand side are
defined in the $\text{MS}$ scheme~\cite{Collins:1984xc}.\footnote{The
renormalization in the $\text{MS}$ scheme to the one-loop order is summarized
in~Appendix~\ref{sec:A}.} Throughout the present paper, we always assume that
the vacuum expectation value is subtracted in renormalized operators.
In~Eq.~\eqref{eq:(2.11)}, the renormalization group functions $\beta$
and~$\gamma_m$ are defined by
\begin{align}
   \beta&\equiv
   \left(\mu\frac{\partial}{\partial\mu}\right)_0g
   =-\frac{1}{2}g\left(\mu\frac{\partial}{\partial\mu}\right)_0\ln Z,
\label{eq:(2.12)}\\
   \gamma_m&\equiv
   -\left(\mu\frac{\partial}{\partial\mu}\right)_0\ln m
   =-\beta\frac{\partial}{\partial g}\ln Z_m,
\label{eq:(2.13)}
\end{align}
where $g$ and~$m$ are the renormalized gauge coupling and the renormalized
mass, respectively, and the derivative with respect to the renormalization
scale~$\mu$ is taken with all bare quantities kept fixed. The
renormalization constants are defined by
\begin{equation}
   g_0^2=\mu^{2\epsilon}g^2Z,\qquad m_0=mZ_m^{-1}.
\label{eq:(2.14)}
\end{equation}
The first few terms of the perturbative expansion of those renormalization
functions read
\begin{align}
   \beta=-b_0g^3-b_1g^5+O(g^7),\qquad\gamma_m=d_0g^2+d_1g^4+O(g^6),
\label{eq:(2.15)}
\end{align}
where~\cite{Caswell:1974gg,Jones:1974mm}
\begin{align}
   b_0&=\frac{1}{(4\pi)^2}
   \left[\frac{11}{3}C_2(G)-\frac{4}{3}T(R)N_{\mathrm{f}}\right],
\label{eq:(2.16)}\\
   b_1&=\frac{1}{(4\pi)^4}
   \left\{
   \frac{34}{3}C_2(G)^2-\left[4C_2(R)+\frac{20}{3}C_2(G)\right]T(R)N_{\mathrm{f}}
   \right\},
\label{eq:(2.17)}
\end{align}
and~\cite{Tarrach:1980up,Nachtmann:1981zg}
\begin{align}
   d_0&=\frac{1}{(4\pi)^2}6C_2(R),
\label{eq:(2.18)}\\
   d_1&=\frac{1}{(4\pi)^4}
   \left\{
   3C_2(R)^2
   +\left[\frac{97}{3}C_2(G)-\frac{20}{3}T(R)N_{\mathrm{f}}\right]C_2(R)
   \right\}.
\label{eq:(2.19)}
\end{align}

\section{Yang--Mills gradient flow}
\label{sec:3}
\subsection{Flow equations and the perturbative expansion}
The Yang--Mills gradient flow is a deformation of a gauge field configuration
generated by a gradient flow in which the Yang--Mills action integral is
regarded as a potential height. To be explicit, for the gauge
potential~$A_\mu(x)$, the flow is defined
by~\cite{Luscher:2010iy,Luscher:2011bx}
\begin{equation}
   \partial_tB_\mu(t,x)=D_\nu G_{\nu\mu}(t,x)
   +\alpha_0D_\mu\partial_\nu B_\nu(t,x),\qquad
   B_\mu(t=0,x)=A_\mu(x),
\label{eq:(3.1)}
\end{equation}
where $t$~is the flow time and $G_{\mu\nu}(t,x)$ is the field strength of the
flowed field,
\begin{equation}
   G_{\mu\nu}(t,x)
   =\partial_\mu B_\nu(t,x)-\partial_\nu B_\mu(t,x)
   +[B_\mu(t,x),B_\nu(t,x)],
\label{eq:(3.2)}
\end{equation}
and the covariant derivative on the gauge field is
\begin{equation}
   D_\mu=\partial_\mu+[B_\mu,\cdot].
\label{eq:(3.3)}
\end{equation}
The first term in the right-hand side of~Eq.~\eqref{eq:(3.1)} is the
``gradient'' in the functional space, $-g_0^2\delta S/\delta B_\mu(t,x)$, where
$S$~is the Yang--Mills action integral for the flowed field. Note that since
$D_\nu G_{\nu\mu}(t,x)=\Delta B_\mu(t,x)+O(B^2)$, Eq.~\eqref{eq:(3.1)} is a sort
of diffusion equation and the flow for~$t>0$ effectively suppresses
high-frequency modes in the configuration. The second term in the right-hand
side of~Eq.~\eqref{eq:(3.1)} with the parameter~$\alpha_0$ is a ``gauge-fixing
term'' that makes the perturbative expansion well defined. It can be shown,
however, that any gauge-invariant quantities are independent of~$\alpha_0$. In
actual perturbative calculation in the next section, we adopt the ``Feynman
gauge'' $\alpha_0=1$ with which the expressions become simplest.

The formal solution of~Eq.~\eqref{eq:(3.1)} is given
by~\cite{Luscher:2010iy,Luscher:2011bx}
\begin{equation}
   B_\mu(t,x)
   =\int\mathrm{d}^Dy\left[
   K_t(x-y)_{\mu\nu}A_\nu(y)
   +\int_0^t\mathrm{d}s\,K_{t-s}(x-y)_{\mu\nu}R_\nu(s,y)
   \right],
\label{eq:(3.4)}
\end{equation}
where\footnote{Throughout the present paper, we use the abbreviation,
\begin{equation}
   \int_p\equiv\int\frac{\mathrm{d}^Dp}{(2\pi)^D}.
\label{eq:(3.5)}
\end{equation}
}
\begin{equation}
   K_t(x)_{\mu\nu}=\int_p\frac{\mathrm{e}^{ipx}}{p^2}
   \left[(\delta_{\mu\nu}p^2-p_\mu p_\nu)\mathrm{e}^{-tp^2}
   +p_\mu p_\nu \mathrm{e}^{-\alpha_0tp^2}\right]
\label{eq:(3.6)}
\end{equation}
is the heat kernel and
\begin{equation}
   R_\mu=2[B_\nu,\partial_\nu B_\mu]
   -[B_\nu,\partial_\mu B_\nu]
   +(\alpha_0-1)[B_\mu,\partial_\nu B_\nu]
   +[B_\nu,[B_\nu,B_\mu]]
\label{eq:(3.7)}
\end{equation}
denotes non-linear interaction terms. By iteratively solving
Eq.~\eqref{eq:(3.4)}, we have a perturbative expansion for the flowed
field~$B_\mu(t,x)$ in terms of the initial value~$A_\nu(y)$.

A similar flow may also be considered for fermion
fields~\cite{Luscher:2013cpa}. For our purpose, it is not necessary that the
flow of fermion fields is a gradient flow of the original fermion action. A
possible choice introduced in~Ref.~\cite{Luscher:2013cpa} is
\begin{align}
   &\partial_t\chi(t,x)=\left[\Delta-\alpha_0\partial_\mu B_\mu(t,x)\right]
   \chi(t,x),\qquad
   \chi(t=0,x)=\psi(x),
\label{eq:(3.8)}\\
   &\partial_t\Bar{\chi}(t,x)
   =\Bar{\chi}(t,x)
   \left[\overleftarrow{\Delta}
   +\alpha_0\partial_\mu B_\mu(t,x)\right],
   \qquad\Bar{\chi}(t=0,x)=\Bar{\psi}(x),
\label{eq:(3.9)}
\end{align}
where
\begin{align}
   &\Delta=D_\mu D_\mu,\qquad D_\mu=\partial_\mu+B_\mu,
\label{eq:(3.10)}\\
   &\overleftarrow{\Delta}=\overleftarrow{D}_\mu\overleftarrow{D}_\mu,
   \qquad\overleftarrow{D}_\mu\equiv\overleftarrow{\partial}_\mu-B_\mu.
\label{eq:(3.11)}
\end{align}
The formal solutions of the above flow equations are
\begin{align}
   \chi(t,x)&=\int\mathrm{d}^Dy\,
   \left[
   K_t(x-y)\psi(y)
   +\int_0^t\mathrm{d}s\,K_{t-s}(x-y)\Delta'\chi(s,y)
   \right],
\label{eq:(3.12)}\\
   \Bar{\chi}(t,x)&=\int\mathrm{d}^Dy\,
   \left[
   \Bar{\psi}(y)K_t(x-y)
   +\int_0^t\mathrm{d}s\,\Bar{\chi}(s,y)\overleftarrow{\Delta}'K_{t-s}(x-y)
   \right],
\label{eq:(3.13)}
\end{align}
where
\begin{equation}
   K_t(x)\equiv\int_p \mathrm{e}^{ipx}\mathrm{e}^{-tp^2}=\frac{\mathrm{e}^{-x^2/4t}}{(4\pi t)^{D/2}},
\label{eq:(3.14)}
\end{equation}
and
\begin{align}
   \Delta'&\equiv(1-\alpha_0)\partial_\mu B_\mu
   +2B_\mu\partial_\mu+B_\mu B_\mu,
\label{eq:(3.15)}\\
   \overleftarrow{\Delta}'&
   \equiv-(1-\alpha_0)\partial_\mu B_\mu
   -2\overleftarrow{\partial}_\mu B_\mu+B_\mu B_\mu.
\label{eq:(3.16)}
\end{align}

The initial values for the above flow, $A_\mu(x)$, $\psi(x)$,
and~$\Bar{\psi}(x)$, are quantum fields being subject to the functional
integral. The quantum correlation functions of the flowed fields are thus
obtained by expressing the flowed fields in terms of original un-flowed fields
(the initial values) and taking the functional average of the latter.
Equations~\eqref{eq:(3.4)}, \eqref{eq:(3.12)}, and~\eqref{eq:(3.13)} provide an
explicit method to carry this out. For example, in the lowest (tree-level)
approximation, we have
\begin{equation}
   \left\langle B_\mu^a(t,x)B_\nu^b(s,y)\right\rangle
   =g_0^2\delta^{ab}\delta_{\mu\nu}\int_p\mathrm{e}^{ip(x-y)}\frac{\mathrm{e}^{-(t+s)p^2}}{p^2},
\label{eq:(3.17)}
\end{equation}
in the ``Feynman gauge'' in which $\lambda_0=\alpha_0=1$, where $\lambda_0$ is
the conventional gauge-fixing parameter. Similarly, for the fermion field, in
the tree-level approximation,
\begin{equation}
   \left\langle\chi(t,x)\Bar{\chi}(s,y)\right\rangle
   =\int_p\mathrm{e}^{ip(x-y)}\frac{\mathrm{e}^{-(t+s)p^2}}{i\Slash{p}+m_0}.
\label{eq:(3.18)}
\end{equation}
Besides these ``quantum propagators,'' we also have heat kernels,
Eqs.~\eqref{eq:(3.6)} and~\eqref{eq:(3.14)}, in the perturbative expansion of
Eqs.~\eqref{eq:(3.4)}, \eqref{eq:(3.12)}, and~\eqref{eq:(3.13)}.

We now explain a diagrammatic representation of the perturbative expansion of
flowed fields (the flow Feynman diagram). For quantum
propagators~\eqref{eq:(3.17)} and~\eqref{eq:(3.18)}, we use the standard
convention that the free propagator of the gauge boson is denoted by a wavy
line and the free propagator of the fermion is denoted by an arrowed straight
line. We stick to these conventions because these are quite natural in a system
containing fermions. In~Refs.~\cite{Suzuki:2013gza} and~\cite{Luscher:2011bx},
on the other hand, the arrowed straight line was adopted to represent the gauge
boson heat kernel~\eqref{eq:(3.6)}. Since we already used this in this paper
for the fermion propagator, we instead use ``doubled lines'' to represent heat
kernels~\eqref{eq:(3.6)} and~\eqref{eq:(3.14)}. For instance,
Figs.~\ref{fig:1}--\ref{fig:5} are one-loop flow Feynman diagrams which
contribute to the two-point function of the flowed fermion field. In these
figures, the doubled straight line represents the fermion heat
kernel~\eqref{eq:(3.14)}; the arrow denotes the flow of the fermion number, not
the direction of the flow time. Similarly, in~Fig.~\ref{fig:15}, the doubled
wavy line is the gauge boson heat kernel~\eqref{eq:(3.6)}. In the present
representation, we thus lose the information of the direction of the flow time,
which is represented by an arrow in~Refs.~\cite{Suzuki:2013gza}
and~\cite{Luscher:2011bx}. This information, however, can readily be traced
back.

Another element of the flow Feynman diagram is the vertex. The vertices that
come from the original action~\eqref{eq:(2.1)} are denoted by filled circles,
while vertices that come through the flow equations, Eqs.~\eqref{eq:(3.7)},
\eqref{eq:(3.15)}, and~\eqref{eq:(3.16)}, are denoted by open circles as
in~Figs.~\ref{fig:1}--\ref{fig:5}; these conventions for the vertex are
identical to those of~Refs.~\cite{Suzuki:2013gza} and~\cite{Luscher:2011bx}.

\subsection{Ringed fermion fields}
A salient feature of the flowed fields is the UV finiteness: Any correlation
functions of the flowed gauge field $B_\mu(t,x)$ with strictly positive~$t$
are, when expressed in terms of renormalized parameters, UV finite
\emph{without\/} the multiplicative (wave function)
renormalization~\cite{Luscher:2011bx}. Moreover, this finiteness persists even
for the equal-point limit. Thus, any correlation functions of any local
products of~$B_\mu(t,x)$ (with $t>0$) are UV finite without further
renormalization other than the parameter renormalization. In other words,
although those local products are given by a certain combination of the bare
gauge field~$A_\mu(x)$ through the flow equation, they are renormalized finite
quantities. A basic reason for this UV finiteness is that the propagator of the
flowed gauge field~\eqref{eq:(3.17)} contains the Gaussian dumping
factor~$\sim \mathrm{e}^{-tp^2}$ which effectively provides a UV cutoff for~$t>0$. To
prove the above finiteness, however, one has to also utilize a
Becchi--Rouet--Stora (BRS) symmetry underlying the present system that is
\emph{inhomogeneous\/} with respect to the gauge
potential~\cite{Luscher:2011bx}.

\begin{figure}
\begin{minipage}{0.3\hsize}
\begin{center}
\includegraphics[width=3cm,clip]{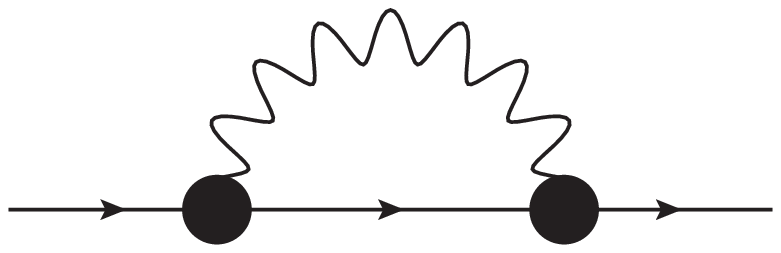}
\caption{C01}
\label{fig:1}
\end{center}
\end{minipage}
\begin{minipage}{0.3\hsize}
\begin{center}
\includegraphics[width=3cm,clip]{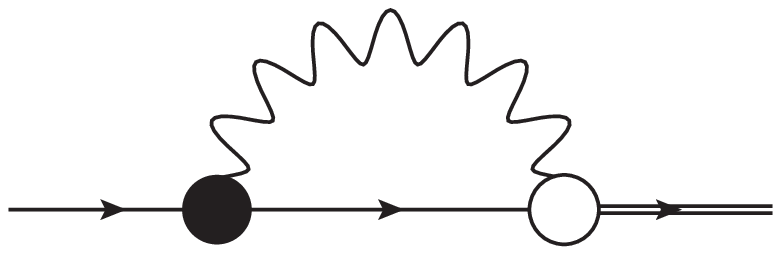}
\caption{C02}
\label{fig:2}
\end{center}
\end{minipage}
\begin{minipage}{0.3\hsize}
\begin{center}
\includegraphics[width=3cm,clip]{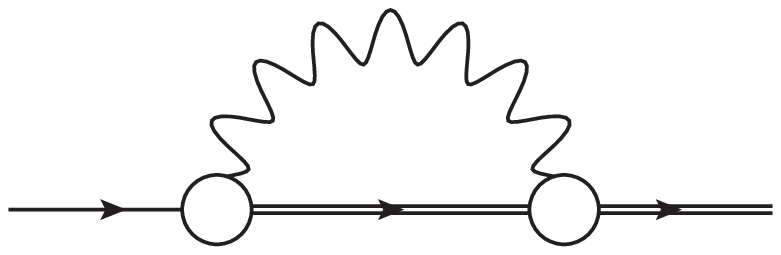}
\caption{C03}
\label{fig:3}
\end{center}
\end{minipage}
\end{figure}

\begin{figure}
\begin{minipage}{0.3\hsize}
\begin{center}
\includegraphics[width=3cm,clip]{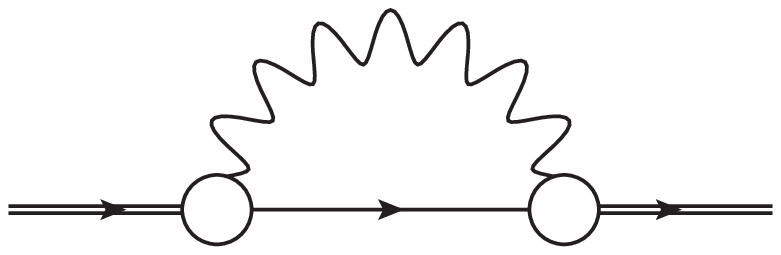}
\caption{C04}
\label{fig:4}
\end{center}
\end{minipage}
\begin{minipage}{0.3\hsize}
\begin{center}
\includegraphics[width=3cm,clip]{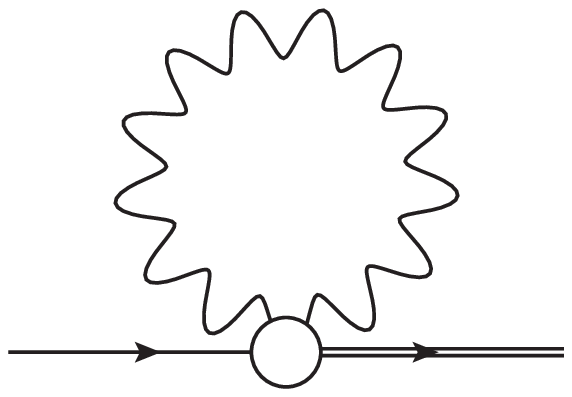}
\caption{C05}
\label{fig:5}
\end{center}
\end{minipage}
\end{figure}

Regrettably, the above finiteness in the first sense does not hold for the
flowed fermion field. It \emph{requires\/} the wave function renormalization.
Although its propagator~\eqref{eq:(3.18)} also possesses the dumping
factor~$\sim \mathrm{e}^{-tp^2}$, the BRS transformation is \emph{homogeneous\/} on the
fermion field (and on general matter fields) and the finiteness proof
in~Ref.~\cite{Luscher:2011bx} does not apply. In fact, computation of the
one-loop diagrams in~Figs.~\ref{fig:1}--\ref{fig:5} (and diagrams with opposite
arrows) shows that the wave function renormalization
\begin{equation}
   \chi_R(t,x)=Z_\chi^{1/2}\chi(t,x),\qquad
   \Bar{\chi}_R(t,x)=Z_\chi^{1/2}\Bar{\chi}(t,x),\qquad
   Z_\chi=1+\frac{g^2}{(4\pi)^2}C_2(R)3\frac{1}{\epsilon}+O(g^4)
\label{eq:(3.19)}
\end{equation}
makes correlation functions UV finite~\cite{Luscher:2013cpa}. Finiteness in the
above second sense still holds: Any correlation functions of any
local products of $\chi_R(t,x)$ and~$\Bar{\chi}_R(t,x)$ remain UV
finite~\cite{Luscher:2013cpa}.

Although the finiteness in the second sense is quite useful for our purpose, we
still need to incorporate the renormalization factor~$Z_\chi$
in~Eq.~\eqref{eq:(3.19)}. This introduces a complication to our problem,
because we have to find a matching factor between $Z_\chi$ in the dimensional
regularization and that in the lattice regularization.

One possible way to avoid this complication is to normalize the fermion fields
by the vacuum expectation value of the fermion kinetic operator:\footnote{In
the kinetic operator, the summation over $N_{\mathrm{f}}$ fermion flavors is understood.
In the first version of the present paper, we used the scalar condensation to
normalize the fermion fields. This choice causes another complication
associated with the massless fermion and the use of the kinetic operator seems
much more appropriate.}
\begin{align}
   \mathring{\chi}(t,x)
   &=\sqrt{\frac{-2\dim(R)N_{\mathrm{f}}}
   {(4\pi)^2t^2
   \left\langle\Bar{\chi}(t,x)\overleftrightarrow{\Slash{D}}\chi(t,x)
   \right\rangle}}
   \,\chi(t,x),
\label{eq:(3.20)}\\
   \mathring{\Bar{\chi}}(t,x)
   &=\sqrt{\frac{-2\dim(R)N_{\mathrm{f}}}
   {(4\pi)^2t^2
   \left\langle\Bar{\chi}(t,x)\overleftrightarrow{\Slash{D}}\chi(t,x)
   \right\rangle}}
   \,\Bar{\chi}(t,x),
\label{eq:(3.21)}
\end{align}
where
\begin{equation}
   \overleftrightarrow{D}_\mu\equiv D_\mu-\overleftarrow{D}_\mu,
\label{eq:(3.22)}
\end{equation}
so that the multiplicative renormalization factor~$Z_\chi$ is cancelled out in
the new ringed variables. Note that the mass dimension
of~$\mathring{\chi}(t,x)$ and~$\mathring{\Bar{\chi}}(t,x)$ is $3/2$ for
any~$D$, while that of~$\chi(t,x)$ and~$\Bar{\chi}(t,x)$ is~$(D-1)/2$.

\begin{figure}
\begin{center}
\includegraphics[width=3cm,clip]{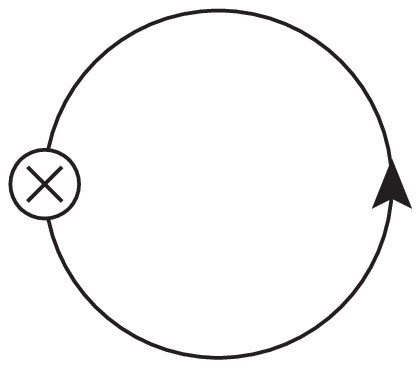}
\caption{D01}
\label{fig:6}
\end{center}
\end{figure}

The vacuum expectation value of the kinetic operator in the lowest (one-loop)
order approximation is given by diagram~D01 in~Fig.~\ref{fig:6}. For
$D=4-2\epsilon$ dimensions, we have
\begin{equation}
   \left\langle\Bar{\chi}(t,x)\overleftrightarrow{\Slash{D}}\chi(t,x)
   \right\rangle
   =\frac{-2\dim(R)N_{\mathrm{f}}}{(4\pi)^2t^2}(8\pi t)^\epsilon\left[1+O(m_0^2t)\right].
\label{eq:(3.23)}
\end{equation}
Note that the mass scale, which is required for the vacuum expectation value,
is supplied by the flow time~$t$ in the present setup. Having obtained this
expression, the constant factors in~Eqs.~\eqref{eq:(3.20)}
and~\eqref{eq:(3.21)} have been chosen so that the difference between the
ringed and the original variables becomes~$O(g_0^2)$ for sufficiently small
flow time ($m_0^2t\ll1$).

\begin{figure}
\begin{minipage}{0.3\hsize}
\begin{center}
\includegraphics[width=3cm,clip]{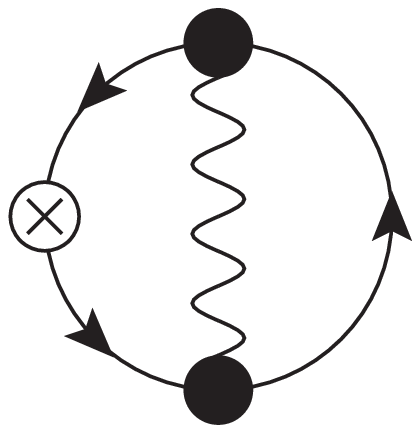}
\caption{D02}
\label{fig:7}
\end{center}
\end{minipage}
\begin{minipage}{0.3\hsize}
\begin{center}
\includegraphics[width=3cm,clip]{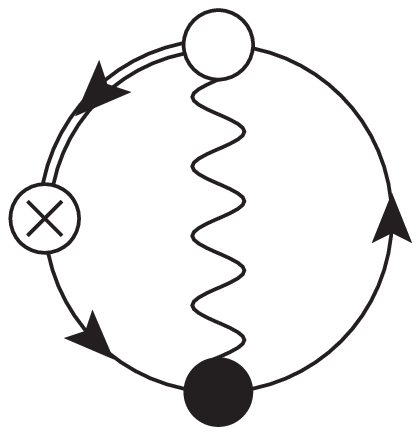}
\caption{D03}
\label{fig:8}
\end{center}
\end{minipage}
\begin{minipage}{0.3\hsize}
\begin{center}
\includegraphics[width=3cm,clip]{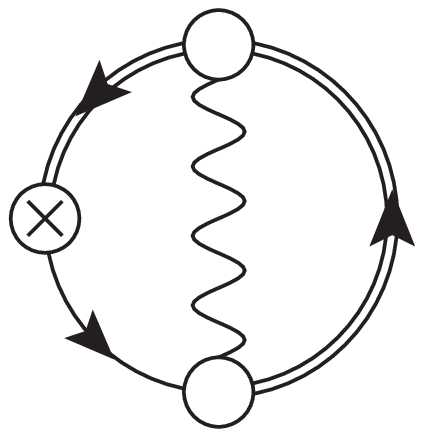}
\caption{D04}
\label{fig:9}
\end{center}
\end{minipage}
\end{figure}

\begin{figure}
\begin{minipage}{0.3\hsize}
\begin{center}
\includegraphics[width=3cm,clip]{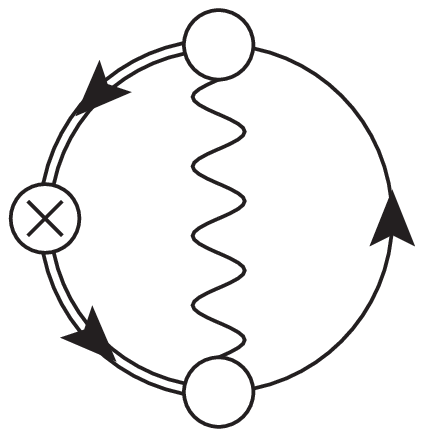}
\caption{D05}
\label{fig:10}
\end{center}
\end{minipage}
\begin{minipage}{0.3\hsize}
\begin{center}
\includegraphics[width=3cm,clip]{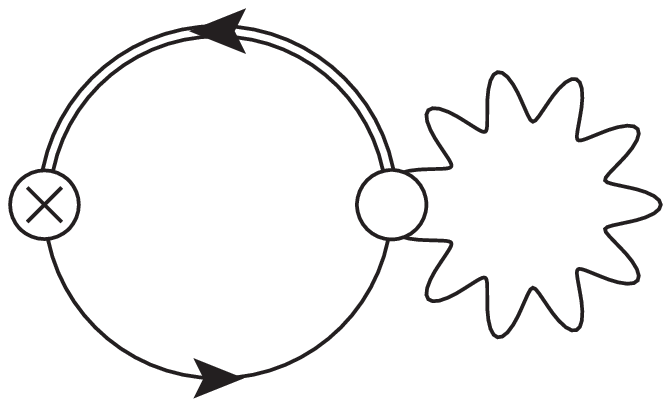}
\caption{D06}
\label{fig:11}
\end{center}
\end{minipage}
\begin{minipage}{0.3\hsize}
\begin{center}
\includegraphics[width=3cm,clip]{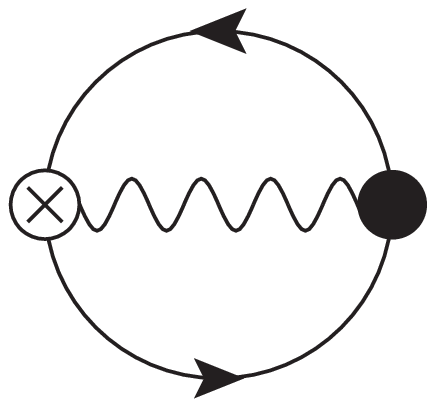}
\caption{D07}
\label{fig:12}
\end{center}
\end{minipage}
\end{figure}

\begin{figure}
\begin{center}
\includegraphics[width=3cm,clip]{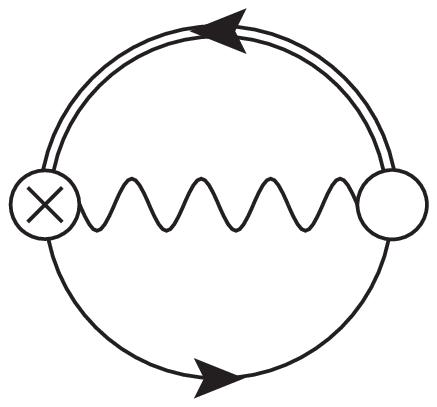}
\caption{D08}
\label{fig:13}
\end{center}
\end{figure}

\begin{table}
\caption{Contribution of each diagram to Eq.~\eqref{eq:(3.24)} in units
of~$\frac{-2\dim(R)N_{\mathrm{f}}}{(4\pi)^2t^2}\frac{g_0^2}{(4\pi)^2}C_2(R)$.}
\label{table:1}
\begin{center}
\renewcommand{\arraystretch}{2.2}
\setlength{\tabcolsep}{20pt}
\begin{tabular}{cr}
\toprule
 diagram & \\
\midrule
 D02 & $-\dfrac{1}{\epsilon}-2\ln(8\pi t)+O(m_0^2t)$ \\
 D03 & $2\dfrac{1}{\epsilon}+4\ln(8\pi t)+2+4\ln2-2\ln3+O(m_0^2t)$ \\
 D04 & $-20\ln2+16\ln3+O(m_0^2t)$ \\
 D05 & $12\ln2-5\ln3+O(m_0^2t)$ \\
 D06 & $-4\dfrac{1}{\epsilon}-8\ln(8\pi t)-2+O(m_0^2t)$ \\
 D07 & $8\ln2-4\ln3+O(m_0^2t)$ \\
 D08 & $-2\ln3+O(m_0^2t)$ \\
\bottomrule
\end{tabular}
\end{center}
\end{table}

The next-to-leading-order (i.e., two-loop) expression for the expectation
value~\eqref{eq:(3.23)} is given by the flow Feynman diagrams
in~Figs.~\ref{fig:7}--\ref{fig:13} (and diagrams with arrows with the opposite
direction). The computation of these diagrams is somewhat complicated but can
be completed in a similar manner to the calculation in~Appendix~B
of~Ref.~\cite{Luscher:2010iy} [with the integration formulas in our
Appendix~\ref{sec:B}, Eqs.~\eqref{eq:(B1)} and~\eqref{eq:(B2)}]. The
contribution of each diagram is tabulated in~Table~\ref{table:1}. In total, we
have
\begin{align}
   &\left\langle\Bar{\chi}(t,x)\overleftrightarrow{\Slash{D}}\chi(t,x)
   \right\rangle
\notag\\
   &=\frac{-2\dim(R)N_{\mathrm{f}}}{(4\pi)^2t^2}
   \left\{(8\pi t)^\epsilon
   +\frac{g_0^2}{(4\pi)^2}C_2(R)
   \left[-3\frac{1}{\epsilon}-6\ln(8\pi t)+\ln(432)\right]
   +O(m_0^2t)+O(g_0^4)\right\}.
\label{eq:(3.24)}
\end{align}
Using Eq.~\eqref{eq:(A1)} for the normalization factor
in~Eqs.~\eqref{eq:(3.20)} and~\eqref{eq:(3.21)}, we have
\begin{equation}
   \frac{-2\dim(R)N_{\mathrm{f}}}
   {(4\pi)^2t^2\left\langle\Bar{\chi}(t,x)
   \overleftrightarrow{\Slash{D}}\chi(t,x)\right\rangle}
   =Z(\epsilon)
   \left\{1+\frac{g^2}{(4\pi)^2}C_2(R)\left[3\frac{1}{\epsilon}-\Phi(t)\right]
   +O(m^2t)+O(g^4)\right\},
\label{eq:(3.25)}
\end{equation}
where
\begin{equation}
   Z(\epsilon)\equiv\frac{1}{(8\pi t)^\epsilon},
\label{eq:(3.26)}
\end{equation}
and
\begin{equation}
   \Phi(t)\equiv-3\ln(8\pi\mu^2t)+\ln(432).
\label{eq:(3.27)}
\end{equation}

\section{Energy--momentum tensor constructed from the flowed
fields}
\label{sec:4}
\subsection{Small flow-time expansion and the renormalization group
argument}
To express the energy--momentum tensor in Eqs.~\eqref{eq:(2.7)}
and~\eqref{eq:(2.10)} in terms of the flowed fields, we consider following
second-rank symmetric tensors (which are even under the CP transformation)
constructed from the flowed fields:
\begin{align}
   \Tilde{\mathcal{O}}_{1\mu\nu}(t,x)&\equiv
   G_{\mu\rho}^a(t,x)G_{\nu\rho}^a(t,x),
\label{eq:(4.1)}\\
   \Tilde{\mathcal{O}}_{2\mu\nu}(t,x)&\equiv
   \delta_{\mu\nu}G_{\rho\sigma}^a(t,x)G_{\rho\sigma}^a(t,x),
\label{eq:(4.2)}\\
   \Tilde{\mathcal{O}}_{3\mu\nu}(t,x)&\equiv
   \mathring{\Bar{\chi}}(t,x)
   \left(\gamma_\mu\overleftrightarrow{D}_\nu
   +\gamma_\nu\overleftrightarrow{D}_\mu\right)
   \mathring{\chi}(t,x),
\label{eq:(4.3)}\\
   \Tilde{\mathcal{O}}_{4\mu\nu}(t,x)&\equiv
   \delta_{\mu\nu}
   \mathring{\Bar{\chi}}(t,x)
   \overleftrightarrow{\Slash{D}}
   \mathring{\chi}(t,x),
\label{eq:(4.4)}\\
   \Tilde{\mathcal{O}}_{5\mu\nu}(t,x)&\equiv
   \delta_{\mu\nu}
   m\mathring{\Bar{\chi}}(t,x)
   \mathring{\chi}(t,x).
\label{eq:(4.5)}
\end{align}
Note that all the above operators $\Tilde{\mathcal{O}}_{i\mu\nu}(t,x)$ are of
dimension~$4$ for any~$D$.

We also introduce corresponding bare operators in the $D$-dimensional
$x$-space:
\begin{align}
   \mathcal{O}_{1\mu\nu}(x)&\equiv
   F_{\mu\rho}^a(x)F_{\nu\rho}^a(x),
\label{eq:(4.6)}\\
   \mathcal{O}_{2\mu\nu}(x)&\equiv
   \delta_{\mu\nu}F_{\rho\sigma}^a(x)F_{\rho\sigma}^a(x),
\label{eq:(4.7)}\\
   \mathcal{O}_{3\mu\nu}(x)&\equiv
   \Bar{\psi}(x)
   \left(\gamma_\mu\overleftrightarrow{D}_\nu
   +\gamma_\nu\overleftrightarrow{D}_\mu\right)
   \psi(x),
\label{eq:(4.8)}\\
   \mathcal{O}_{4\mu\nu}(x)&\equiv
   \delta_{\mu\nu}
   \Bar{\psi}(x)
   \overleftrightarrow{\Slash{D}}
   \psi(x),
\label{eq:(4.9)}\\
   \mathcal{O}_{5\mu\nu}(x)&\equiv
   \delta_{\mu\nu}
   m_0\Bar{\psi}(x)
   \psi(x).
\label{eq:(4.10)}
\end{align}
The mass dimension of $\mathcal{O}_{1\mu\nu}(x)$ and~$\mathcal{O}_{2\mu\nu}(x)$
is~$4$, while that of $\mathcal{O}_{3\mu\nu}(x)$, $\mathcal{O}_{4\mu\nu}(x)$,
and~$\mathcal{O}_{5\mu\nu}(x)$ is~$D$.

We now consider the situation in which the flow time~$t$
in~Eqs.~\eqref{eq:(4.1)}--\eqref{eq:(4.5)} is very small. Since a flowed field
at position~$x$ with a flow time~$t$ is a combination of un-flowed fields in
the vicinity of~$x$ of radius~$\sim\sqrt{8t}$,\footnote{This follows from
the fact that the flow equations are basically the diffusion equation.} the
operators~\eqref{eq:(4.1)}--\eqref{eq:(4.5)} can be regarded as local operators
in $x$-space in the $t\to0$ limit. Since the bare operators
in~Eqs.~\eqref{eq:(4.6)}--\eqref{eq:(4.10)} span a complete set of symmetric
second-rank gauge-invariant local operators of dimension~$4$ (for $D\to4$)
which are even under CP, we have the following (asymptotic) expansion for
small~$t$:
\begin{equation}
   \Tilde{\mathcal{O}}_{i\mu\nu}(t,x)
   =\left\langle\Tilde{\mathcal{O}}_{i\mu\nu}(t,x)\right\rangle
   +\zeta_{ij}(t)
   \left[\mathcal{O}_{j\mu\nu}(x)
   -\left\langle\mathcal{O}_{j\mu\nu}(x)\right\rangle\right]+O(t),
\label{eq:(4.11)}
\end{equation}
where the abbreviated terms are contributions of operators of mass dimension
$6$ (for $D\to4$) or higher.

In writing down the small flow-time expansion~\eqref{eq:(4.11)}, we have
assumed that there is no other $D$-dimensional composite operator at the
point~$x$. That the expansion~\eqref{eq:(4.11)} cannot necessarily hold when
there is another operator at the point~$x$ [say, $\mathcal{P}(x)$] can be seen
by noting that the product of the left-hand side of Eq.~\eqref{eq:(4.11)}
with~$\mathcal{P}(x)$ does not possess any divergence for~$t>0$, while each
term in the right-hand side can have an equal-point singularity
with~$\mathcal{P}(x)$ because the operators in the right-hand side
of~Eq.~\eqref{eq:(4.11)} are $D$-dimensional (i.e., non-flowed) composite
operators. This is a contradiction if Eq.~\eqref{eq:(4.11)} holds. This implies
that the formula we will derive for the energy--momentum tensor below holds
\emph{only when the energy--momentum tensor has no overlap with other
operators}.\footnote{This important point was not fully recognized
in~Ref.~\cite{Suzuki:2013gza}.} In particular, we cannot say anything about
whether the Ward--Takahashi relation~\eqref{eq:(2.9)} is reproduced with our
construction. Still, our construction is expected to have a correct
normalization because it is determined from matching the energy--momentum
tensor in the dimensional regularization which fulfills~Eq.~\eqref{eq:(2.9)}.
Our lattice energy--momentum tensor is thus useful to compute correlation
functions in which the energy--momentum tensor is separated from other
operators. This is the case for correlation functions relevant to the
viscosities~\cite{Nakamura:2004sy,Meyer:2007ic,Meyer:2007dy}, for example.

The expansion~\eqref{eq:(4.11)} may be inverted as
\begin{equation}
   \mathcal{O}_{i\mu\nu}(x)-\left\langle\mathcal{O}_{i\mu\nu}(x)\right\rangle
   =\left(\zeta^{-1}\right)_{ij}(t)
   \left[\Tilde{\mathcal{O}}_{j\mu\nu}(t,x)
   -\left\langle\Tilde{\mathcal{O}}_{j\mu\nu}(t,x)\right\rangle\right]
   +O(t),
\label{eq:(4.12)}
\end{equation}
where $\zeta^{-1}$ denotes the inverse matrix of~$\zeta$. Then, by substituting
this relation into the energy--momentum tensor~\eqref{eq:(2.7)}
and~\eqref{eq:(2.10)} in terms of the bare operators,
\begin{align}
   \left\{T_{\mu\nu}\right\}_R(x)
   &=\frac{1}{g_0^2}\left\{
   \mathcal{O}_{1\mu\nu}(x)
   -\left\langle\mathcal{O}_{1\mu\nu}(x)\right\rangle
   -\frac{1}{4}
   \left[\mathcal{O}_{2\mu\nu}(x)
   -\left\langle\mathcal{O}_{2\mu\nu}(x)\right\rangle\right]
   \right\}
\notag\\
   &\qquad{}
   +\frac{1}{4}
   \left[\mathcal{O}_{3\mu\nu}(x)
   -\left\langle\mathcal{O}_{3\mu\nu}(x)\right\rangle\right]
   -\frac{1}{2}
   \left[\mathcal{O}_{4\mu\nu}(x)
   -\left\langle\mathcal{O}_{4\mu\nu}(x)\right\rangle\right]
\notag\\
   &\qquad\qquad{}
   -\left[\mathcal{O}_{5\mu\nu}(x)
   -\left\langle\mathcal{O}_{5\mu\nu}(x)\right\rangle\right],
\label{eq:(4.13)}
\end{align}
we have the expression (for $D=4$)
\begin{align}
   \left\{T_{\mu\nu}\right\}_R(x)
   &=c_1(t)\left[
   \Tilde{\mathcal{O}}_{1\mu\nu}(t,x)
   -\frac{1}{4}\Tilde{\mathcal{O}}_{2\mu\nu}(t,x)
   \right]
\notag\\
   &\qquad{}
   +c_2(t)\left[
   \Tilde{\mathcal{O}}_{2\mu\nu}(t,x)
   -\left\langle\Tilde{\mathcal{O}}_{2\mu\nu}(t,x)\right\rangle
   \right]
\notag\\
   &\qquad\qquad{}
   +c_3(t)\left[
   \Tilde{\mathcal{O}}_{3\mu\nu}(t,x)
   -2\Tilde{\mathcal{O}}_{4\mu\nu}(t,x)
   -\left\langle
   \Tilde{\mathcal{O}}_{3\mu\nu}(t,x)
   -2\Tilde{\mathcal{O}}_{4\mu\nu}(t,x)
   \right\rangle
   \right]
\notag\\
   &\qquad\qquad\qquad{}
   +c_4(t)\left[
   \Tilde{\mathcal{O}}_{4\mu\nu}(t,x)
   -\left\langle\Tilde{\mathcal{O}}_{4\mu\nu}(t,x)\right\rangle
   \right]
\notag\\
   &\qquad\qquad\qquad\qquad{}
   +c_5(t)\left[
   \Tilde{\mathcal{O}}_{5\mu\nu}(t,x)
   -\left\langle\Tilde{\mathcal{O}}_{5\mu\nu}(t,x)\right\rangle
   \right]+O(t),
\label{eq:(4.14)}
\end{align}
where
\begin{align}
   &c_1(t)=\Tilde{c}_1(t),\qquad
   c_2(t)=\Tilde{c}_2(t)+\frac{1}{4}c_1(t),
\notag\\
   &c_3(t)=\Tilde{c}_3(t),\qquad
   c_4(t)=\Tilde{c}_4(t)+2c_3(t),\qquad
   c_5(t)=\Tilde{c}_5(t),
\label{eq:(4.15)}
\end{align}
and
\begin{equation}
   \Tilde{c}_i(t)\equiv\frac{1}{g_0^2}\left\{
   \left(\zeta^{-1}\right)_{1i}(t)
   -\frac{1}{4}\left(\zeta^{-1}\right)_{2i}(t)\right\}
   +\frac{1}{4}\left(\zeta^{-1}\right)_{3i}(t)
   -\frac{1}{2}\left(\zeta^{-1}\right)_{4i}(t)
   -\left(\zeta^{-1}\right)_{5i}(t).
\label{eq:(4.16)}
\end{equation}
In Eq.~\eqref{eq:(4.14)}, we have used the fact that the finite operator
$\Tilde{\mathcal{O}}_{1\mu\nu}(t,x)-(1/4)\Tilde{\mathcal{O}}_{2\mu\nu}(t,x)$ is
traceless in~$D=4$ and thus has no vacuum expectation value.
Equation~\eqref{eq:(4.14)} shows that if one knows the $t\to0$ behavior of the
coefficients~$c_i(t)$, the energy--momentum tensor can be obtained as the
$t\to0$ limit of the combination in the right-hand side. As already noted,
since the composite operators~\eqref{eq:(4.1)}--\eqref{eq:(4.5)} constructed
from (ringed) flowed fields should be independent of the regularization
adopted, one may use the lattice regularization to compute correlation
functions of the quantity in the right-hand side of~Eq.~\eqref{eq:(4.14)}. This
provides a possible method to compute correlation functions of the
correctly normalized conserved energy--momentum tensor with the lattice
regularization.

Thus, we are interested in the $t\to0$ behavior of the coefficients~$c_i(t)$
in~Eq.~\eqref{eq:(4.14)}. Quite interestingly, one can argue that the
coefficients~$c_i(t)$ can be evaluated by the perturbation theory for~$t\to0$
thanks to the asymptotic freedom. To see this, we apply
\begin{equation}
   \left(\mu\frac{\partial}{\partial\mu}\right)_0
\label{eq:(4.17)}
\end{equation}
on both sides of~Eq.~\eqref{eq:(4.14)}, where $\mu$ is the renormalization
scale and the subscript~$0$ implies that the derivative is taken while all bare
quantities are kept fixed. Since the energy--momentum tensor is not
multiplicatively renormalized,
$(\mu\partial/\partial\mu)_0(\text{left-hand side of Eq.~\eqref{eq:(4.14)}})=0$.
On the right-hand side, since $\Tilde{O}_{1,2,3,4\mu\nu}(t,x)$
and~$(1/m)\Tilde{O}_{5\mu\nu}(t,x)$ are entirely given by bare quantities
through the flow equations, we have
\begin{equation}
   \left(\mu\frac{\partial}{\partial\mu}\right)_0\Tilde{O}_{1,2,3,4\mu\nu}(t,x)
   =\left(\mu\frac{\partial}{\partial\mu}\right)_0
   \frac{1}{m}\Tilde{O}_{5\mu\nu}(t,x)
   =0.
\label{eq:(4.18)}
\end{equation}
These observations imply
\begin{equation}
   \left(\mu\frac{\partial}{\partial\mu}\right)_0c_{1,2,3,4}(t)
   =\left(\mu\frac{\partial}{\partial\mu}\right)_0mc_5(t)
   =0.
\label{eq:(4.19)}
\end{equation}
Then the standard renormalization group argument says that $c_{1,2,3,4}(t)$
and~$mc_5(t)$ are independent of the renormalization scale, if the renormalized
parameters in these quantities are replaced by running parameters defined by
\begin{align}
   &q\frac{\mathrm{d}\Bar{g}(q)}{\mathrm{d}q}=\beta(\Bar{g}(q)),\qquad
   \Bar{g}(q=\mu)=g,
\label{eq:(4.20)}\\
   &q\frac{\mathrm{d}\Bar{m}(q)}{\mathrm{d}q}=-\gamma_m(\Bar{g}(q))\Bar{m}(q),\qquad
   \Bar{m}(q=\mu)=m,
\label{eq:(4.21)}
\end{align}
where $\mu$ is the original renormalization scale. Thus, since $c_{1,2,3,4}(t)$
and~$mc_5(t)$ are independent of the renormalization scale, two possible
choices, $q=\mu$ and~$q=1/\sqrt{8t}$, should give an identical result. In this
way, we infer that
\begin{align}
   c_{1,2,3,4}(t)(g,m;\mu)
   &=c_{1,2,3,4}(t)(\Bar{g}(1/\sqrt{8t}),\Bar{m}(1/\sqrt{8t});1/\sqrt{8t}),
\label{eq:(4.22)}\\
   c_5(t)(g,m;\mu)
   &=\frac{\Bar{m}(1/\sqrt{8t})}{m}
   c_5(t)(\Bar{g}(1/\sqrt{8t}),\Bar{m}(1/\sqrt{8t});1/\sqrt{8t}),
\label{eq:(4.23)}
\end{align}
where we have explicitly written the dependence of $c_i(t)$ on renormalized
parameters and on the renormalization scale. Finally, since the running gauge
coupling $\Bar{g}(1/\sqrt{8t})\to0$ for~$t\to0$ thanks to the asymptotic
freedom, we expect that we can compute $c_i(t)$ for~$t\to0$ by using the
perturbation theory. Although we are interested in low-energy physics for which
the perturbation theory is ineffective, the coefficients~$c_i(t)$
in~Eq.~\eqref{eq:(4.14)} for~$t\to0$ can be evaluated by perturbation theory;
this might be regarded as a sort of factorization.

\subsection{$c_i(t)$ to the one-loop order}
We thus evaluate the above coefficients~$c_i(t)$ in~Eq.~\eqref{eq:(4.14)} to
the one-loop order approximation. For this, we compute the mixing coefficients
$\zeta_{ij}(t)$ in~Eq.~\eqref{eq:(4.11)} to the one-loop order. Then $c_i(t)$
are obtained by~Eqs.~\eqref{eq:(4.15)} and~\eqref{eq:(4.16)}. The loop
expansion of~$\zeta_{ij}(t)$ would yield
\begin{equation}
   \zeta_{ij}(t)=\delta_{ij}+\zeta_{ij}^{(1)}(t)+\zeta_{ij}^{(2)}(t)+\dotsb,
\label{eq:(4.24)}
\end{equation}
for $j=1$ and~$j=2$, where the superscript denotes the loop order, and for
$j=3$, $4$, and~$5$, by taking the factor in~Eq.~\eqref{eq:(3.26)} into
account, we set 
\begin{equation}
   \zeta_{ij}(t)=Z(\epsilon)
   \left[\delta_{ij}+\zeta_{ij}^{(1)}(t)+\zeta_{ij}^{(2)}(t)+\dotsb\right].
\label{eq:(4.25)}
\end{equation}

As Ref.~\cite{Suzuki:2013gza}, it is straightforward to compute
$\zeta_{ij}^{(1)}(t)$.\footnote{The justification for the following
computational prescription was not well explained in previous versions of the
present paper: One may wonder why the one-loop matching
coefficients~$\zeta_{ij}^{(1)}(t)$ can be read off from the correlation
functions~\eqref{eq:(4.26)} and~\eqref{eq:(4.29)} alone, without computing
corresponding correlation functions in which flowed composite
operators~$\Tilde{\mathcal{O}}_{i\mu\nu}(t,x)$ are replaced by bare
ones~$\mathcal{O}_{i\mu\nu}(x)$. The justification is directly related to our
way of treatment of infrared (IR) divergences and we supplement detailed
explanation in~Appendix~\ref{sec:D}. We are quite grateful to a referee of the
present paper for suggestions on this point.} For example, to obtain
$\zeta_{ij}^{(1)}(t)$ with~$j=1$ and~$2$, we consider the correlation function
\begin{equation}
   \left\langle\Tilde{\mathcal{O}}_{i\mu\nu}(t,x)
   A_\beta^b(y)A_\gamma^c(z)\right\rangle.
\label{eq:(4.26)}
\end{equation}
In the Feynman gauge which we use throughout the present paper, this has the
structure
\begin{equation}
   g_0^2\int_k\frac{\mathrm{e}^{ik(x-y)}}{k^2}g_0^2\int_\ell\frac{\mathrm{e}^{i\ell(x-z)}}{\ell^2}
   \delta^{bc}\mathcal{M}_{\mu\nu,\beta\gamma}(k,\ell).
\label{eq:(4.27)}
\end{equation}
After expanding $\mathcal{M}_{\mu\nu,\beta\gamma}(k,\ell)$ to~$O(k,\ell)$, we can
make use of the following correspondence to read off the operator mixing:
\begin{equation}
   ik_\mu i\ell_\nu\delta_{\beta\gamma}\to F_{\mu\rho}^a(x)F_{\nu\rho}^a(x),\qquad
   ik\cdot i\ell\delta_{\mu\nu}\delta_{\beta\gamma}
   \to\frac{1}{4}\delta_{\mu\nu}F_{\rho\sigma}^a(x)F_{\rho\sigma}^a(x).
\label{eq:(4.28)}
\end{equation}
In this way, we obtain~$\zeta_{ij}^{(1)}(t)$ with $j=1$ and~$2$.\footnote{For
the momentum integration, we use the integration formulas
in~Appendix~\ref{sec:B}.}

Similarly, to obtain $\zeta_{ij}^{(1)}(t)$ with $j=3$, $4$, and~$5$, we consider
\begin{equation}
   \left\langle\Tilde{\mathcal{O}}_{i\mu\nu}(t,x)
   \psi(y)\Bar{\psi}(z)\right\rangle,
\label{eq:(4.29)}
\end{equation}
whose general structure reads
\begin{equation}
   \int_k\frac{\mathrm{e}^{ik(y-x)}}{i\Slash{k}+m_0}
   \mathcal{M}_{\mu\nu}(k,\ell)
   \int_\ell\frac{\mathrm{e}^{i\ell(x-z)}}{i\Slash{\ell}+m_0}.
\label{eq:(4.30)}
\end{equation}
We then expand $\mathcal{M}_{\mu\nu}(k,\ell)$ to~$O(k)$ and~$O(\ell)$ and use the
correspondence
\begin{equation}
   \gamma_\mu i(k+\ell)_\nu+\gamma_\nu i(k+\ell)_\mu\to
   \Bar{\psi}(x)\left[
   \gamma_\mu\overleftrightarrow{D}_\nu+\gamma_\nu\overleftrightarrow{D}_\mu
   \right]\psi(x),\qquad
   \delta_{\mu\nu}\to\delta_{\mu\nu}\Bar{\psi}(x)\psi(x),
\label{eq:(4.31)}
\end{equation}
to read off the operator mixing.





\begin{figure}
\begin{minipage}{0.3\hsize}
\begin{center}
\includegraphics[width=3cm,clip]{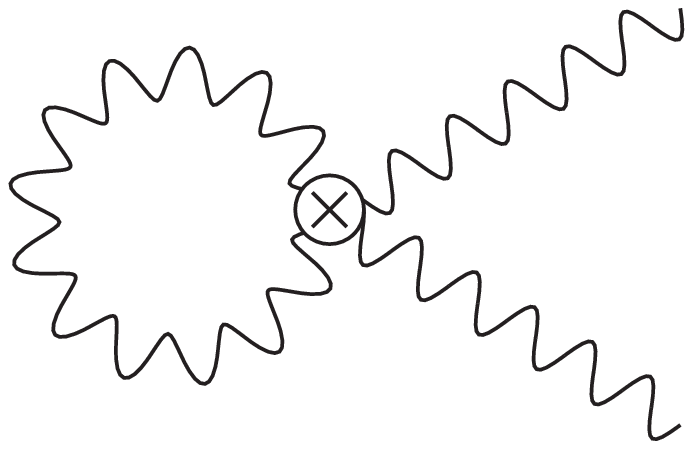}
\caption{A03}
\label{fig:14}
\end{center}
\end{minipage}
\begin{minipage}{0.3\hsize}
\begin{center}
\includegraphics[width=3cm,clip]{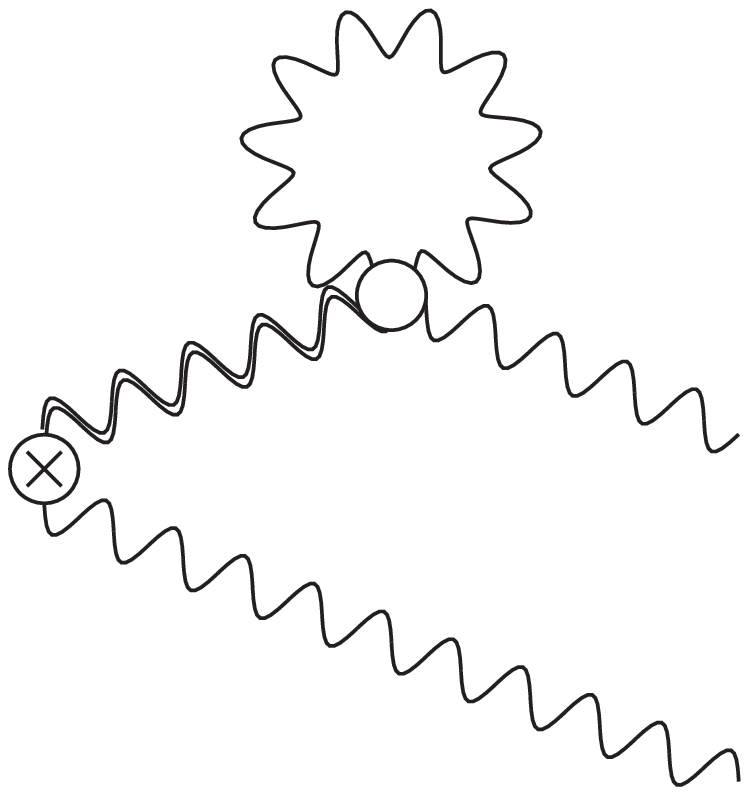}
\caption{A04}
\label{fig:15}
\end{center}
\end{minipage}
\begin{minipage}{0.3\hsize}
\begin{center}
\includegraphics[width=3cm,clip]{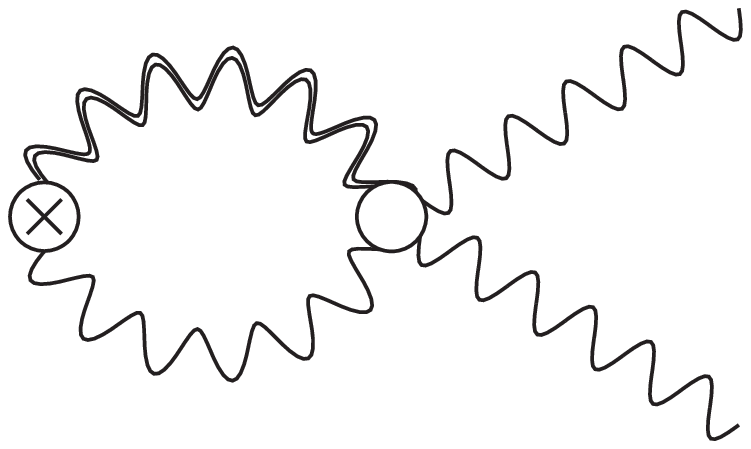}
\caption{A05}
\label{fig:16}
\end{center}
\end{minipage}
\end{figure}

\begin{figure}
\begin{minipage}{0.3\hsize}
\begin{center}
\includegraphics[width=3cm,clip]{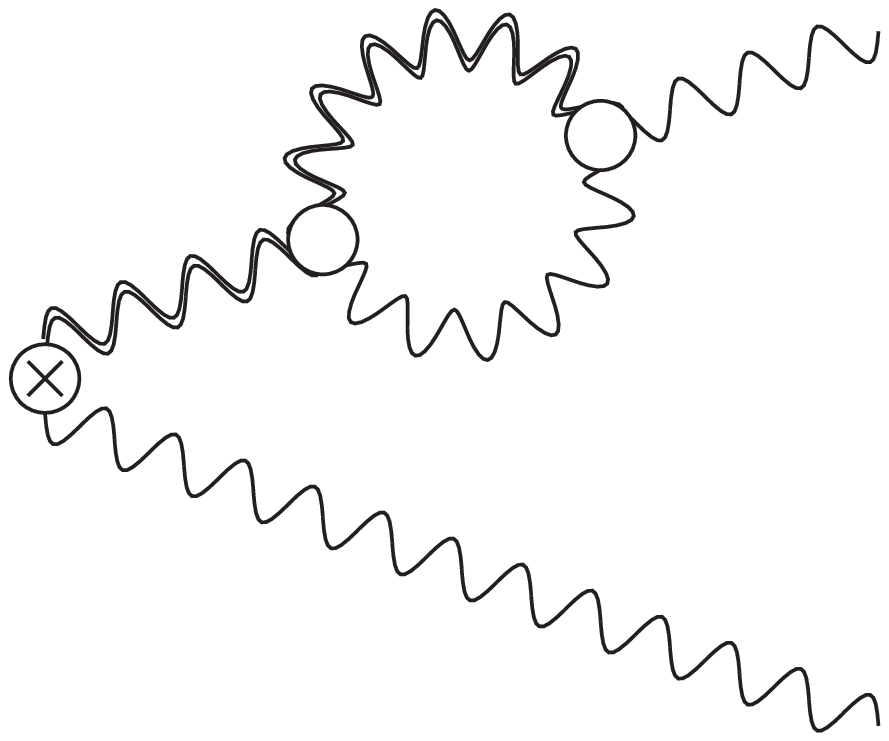}
\caption{A06}
\label{fig:17}
\end{center}
\end{minipage}
\begin{minipage}{0.3\hsize}
\begin{center}
\includegraphics[width=3cm,clip]{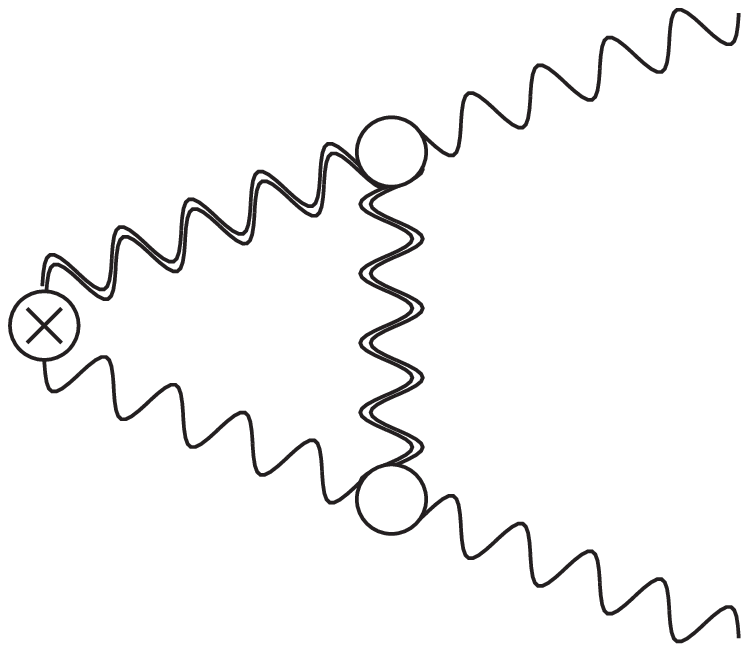}
\caption{A07}
\label{fig:18}
\end{center}
\end{minipage}
\begin{minipage}{0.3\hsize}
\begin{center}
\includegraphics[width=3cm,clip]{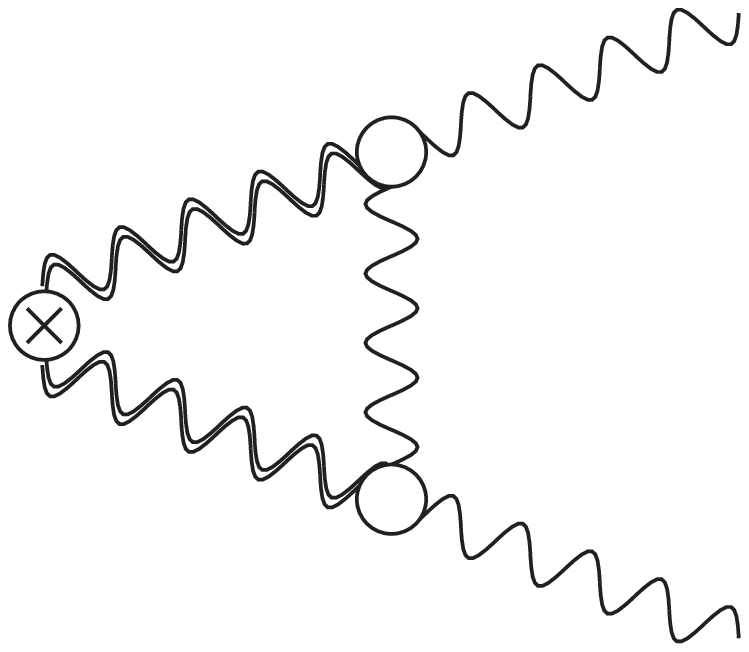}
\caption{A08}
\label{fig:19}
\end{center}
\end{minipage}
\end{figure}

\begin{figure}
\begin{minipage}{0.3\hsize}
\begin{center}
\includegraphics[width=3cm,clip]{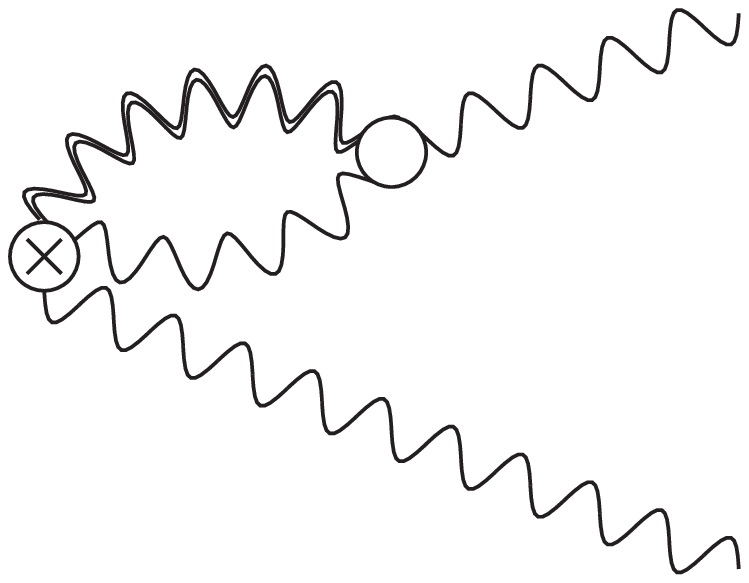}
\caption{A09}
\label{fig:20}
\end{center}
\end{minipage}
\begin{minipage}{0.3\hsize}
\begin{center}
\includegraphics[width=3cm,clip]{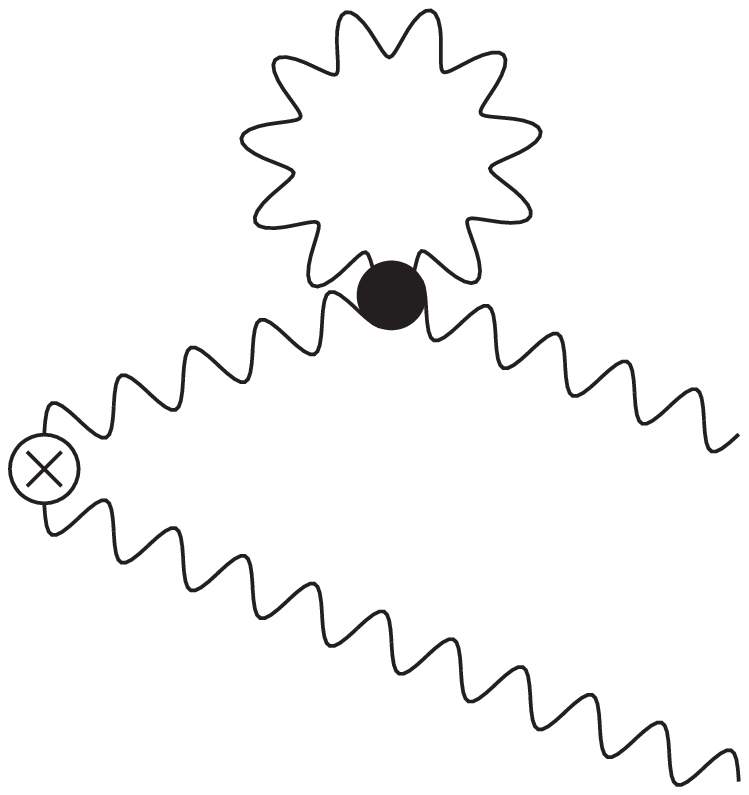}
\caption{A10}
\label{fig:21}
\end{center}
\end{minipage}
\begin{minipage}{0.3\hsize}
\begin{center}
\includegraphics[width=3cm,clip]{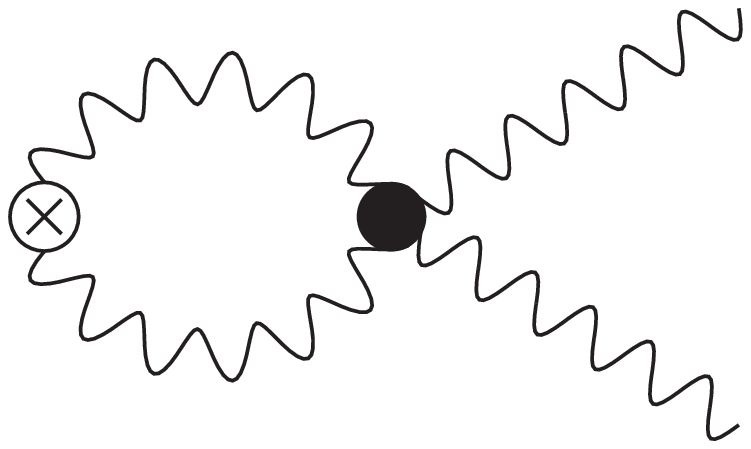}
\caption{A11}
\label{fig:22}
\end{center}
\end{minipage}
\end{figure}

\begin{figure}
\begin{minipage}{0.3\hsize}
\begin{center}
\includegraphics[width=3cm,clip]{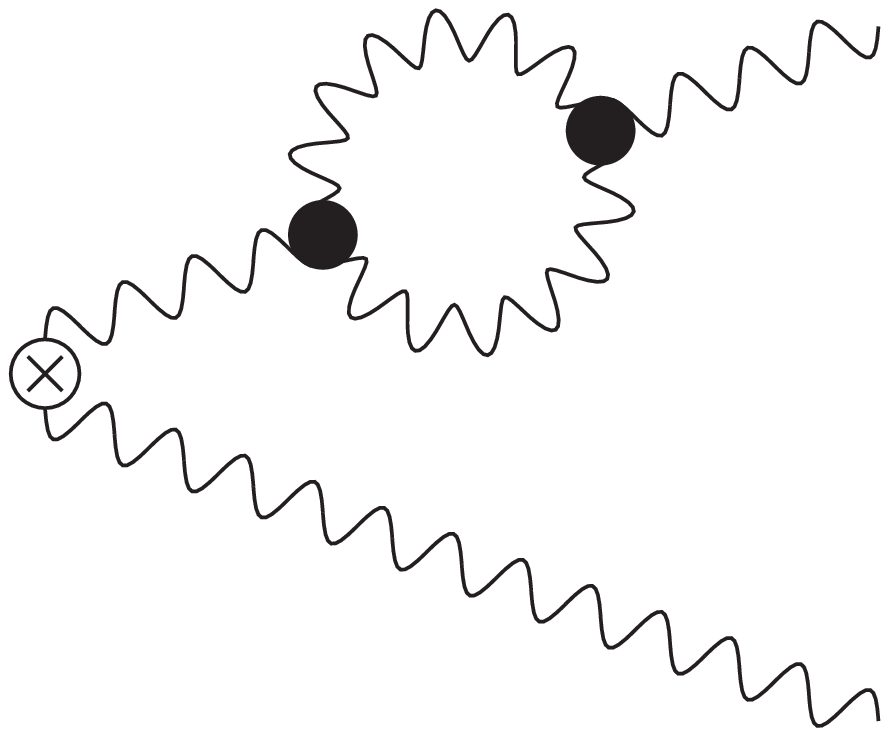}
\caption{A12}
\label{fig:23}
\end{center}
\end{minipage}
\begin{minipage}{0.3\hsize}
\begin{center}
\includegraphics[width=3cm,clip]{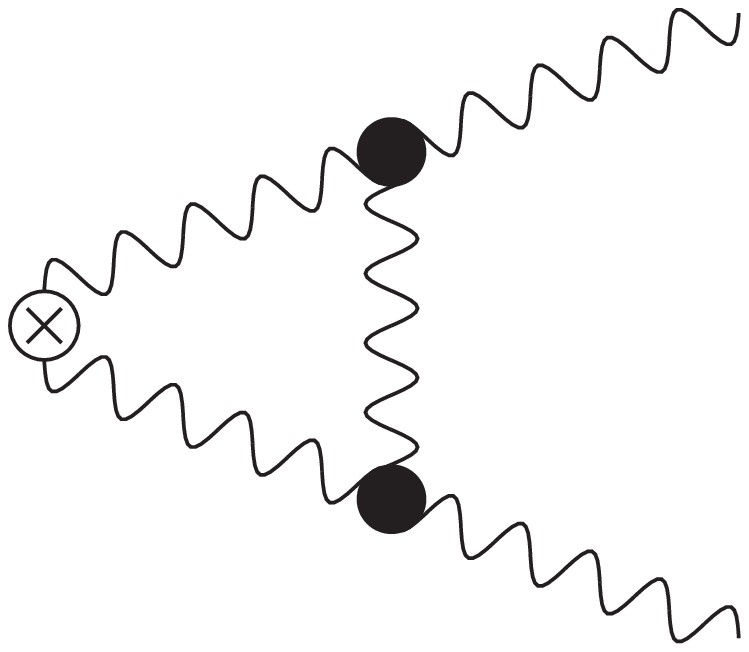}
\caption{A13}
\label{fig:24}
\end{center}
\end{minipage}
\begin{minipage}{0.3\hsize}
\begin{center}
\includegraphics[width=3cm,clip]{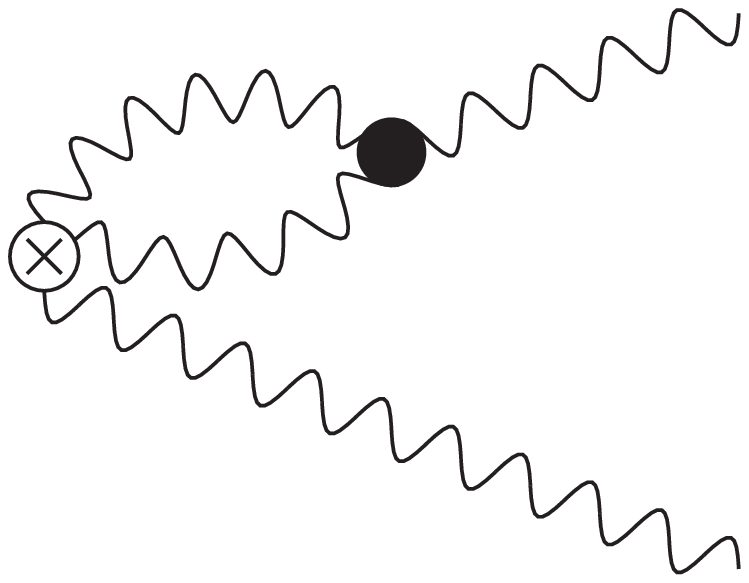}
\caption{A14}
\label{fig:25}
\end{center}
\end{minipage}
\end{figure}

\begin{figure}
\begin{minipage}{0.3\hsize}
\begin{center}
\includegraphics[width=3cm,clip]{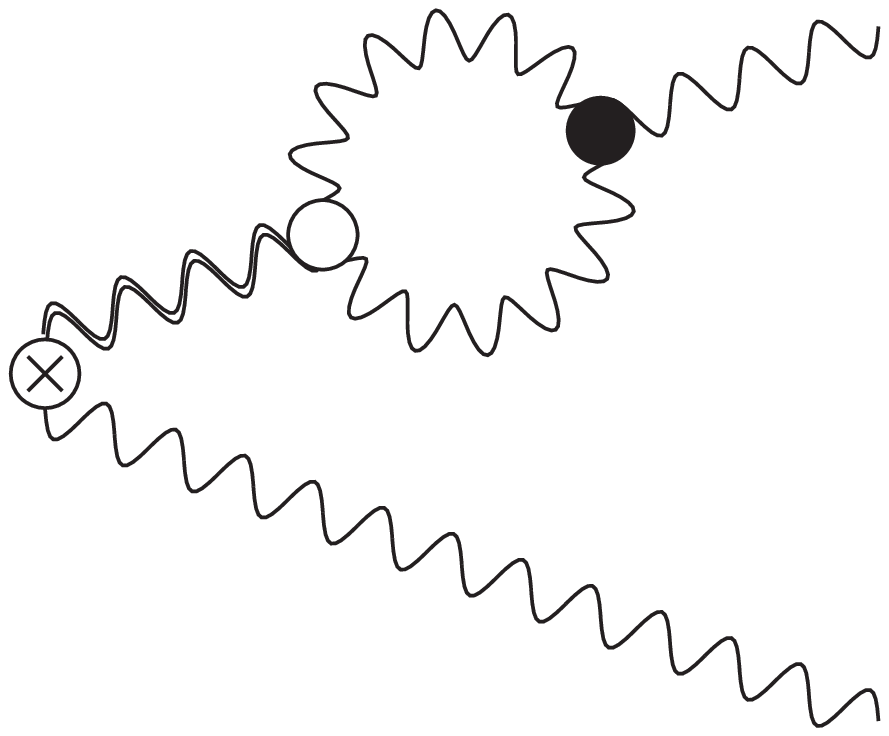}
\caption{A15}
\label{fig:26}
\end{center}
\end{minipage}
\begin{minipage}{0.3\hsize}
\begin{center}
\includegraphics[width=3cm,clip]{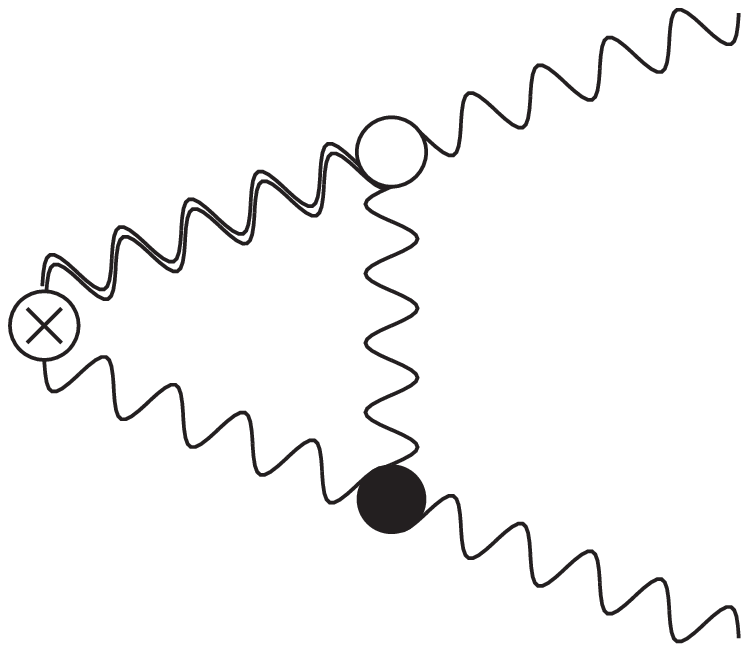}
\caption{A16}
\label{fig:27}
\end{center}
\end{minipage}
\begin{minipage}{0.3\hsize}
\begin{center}
\includegraphics[width=3cm,clip]{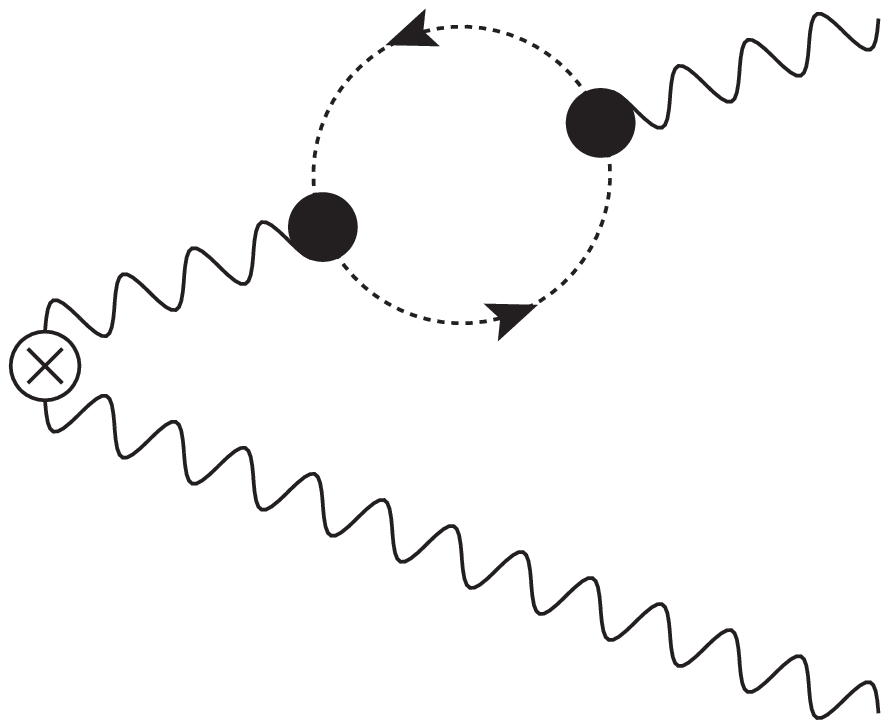}
\caption{A17}
\label{fig:28}
\end{center}
\end{minipage}
\end{figure}

\begin{figure}
\begin{minipage}{0.3\hsize}
\begin{center}
\includegraphics[width=3cm,clip]{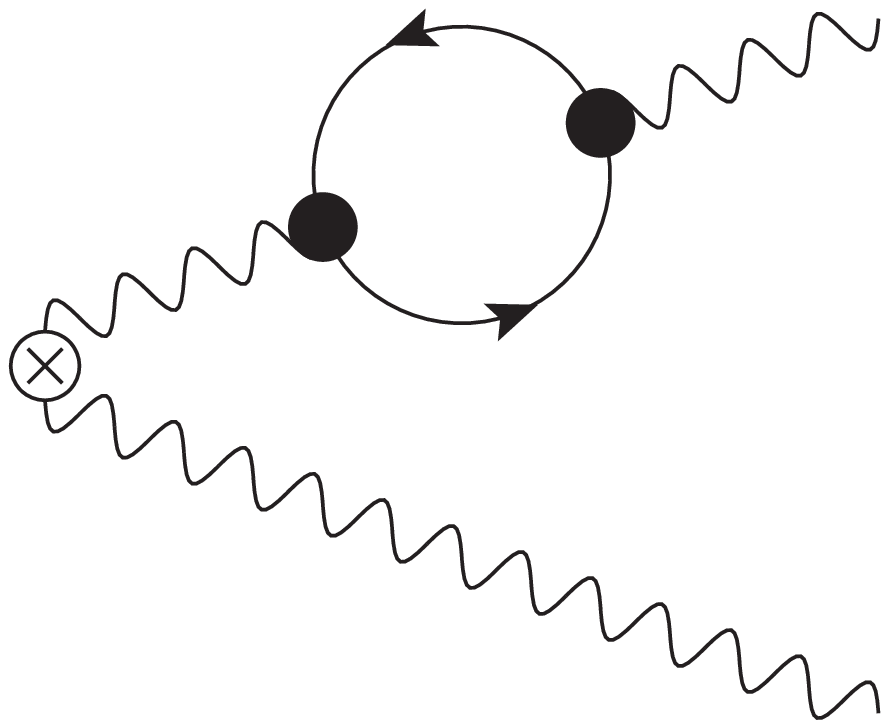}
\caption{A18}
\label{fig:29}
\end{center}
\end{minipage}
\begin{minipage}{0.3\hsize}
\begin{center}
\includegraphics[width=3cm,clip]{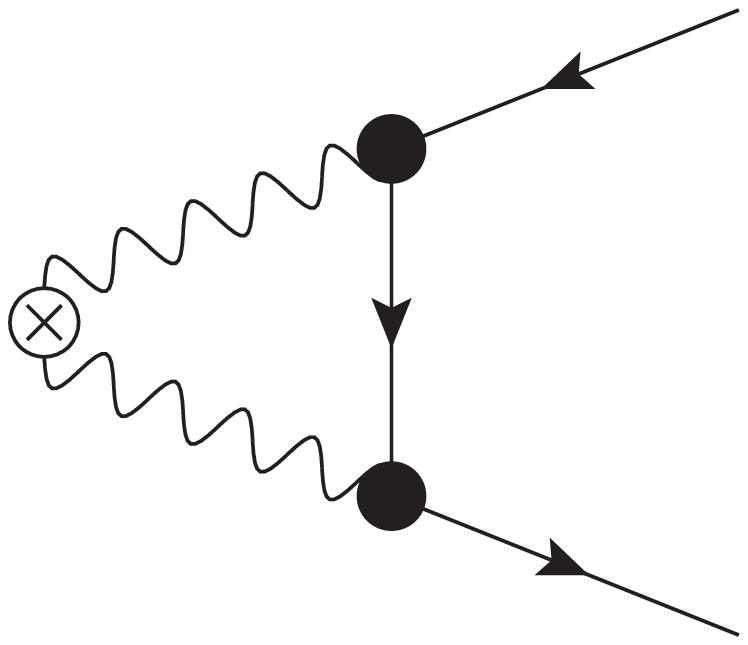}
\caption{A19}
\label{fig:30}
\end{center}
\end{minipage}
\end{figure}
For $\zeta_{1j}^{(1)}(t)$, diagrams A03, A04, A05, A06, A07, A08, A09, A11, A13,
A14, A15, A16, and A19 in~Figs.~\ref{fig:14}--\ref{fig:30}
contribute.\footnote{Diagrams A10, A12, A17 (where the dotted line represents
the ghost propagator), and A18 in these figures correspond to the conventional
wave function renormalization and should be omitted in computing the operator
mixing.} In these and following diagrams, the cross generically represents one
of composite operators in~Eqs.~\eqref{eq:(4.1)}--\eqref{eq:(4.5)} [in the
present case, $\Tilde{\mathcal{O}}_{1\mu\nu}(t,x)$]. Apart from~A19, we can
borrow the results from~Ref.~\cite{Suzuki:2013gza} for these diagrams. For
completeness, these results are reproduced in the present convention
in~Table~\ref{table:0} of~Appendix~\ref{sec:C}. Combined with the contribution
of diagram~A19, we have
\begin{align}
   \zeta_{11}^{(1)}(t)
   &=\frac{g_0^2}{(4\pi)^2}C_2(G)
   \left[\frac{11}{3}\epsilon(t)^{-1}+\frac{7}{3}\right],
\label{eq:(4.32)}\\
   \zeta_{12}^{(1)}(t)
   &=\frac{g_0^2}{(4\pi)^2}C_2(G)
   \left[-\frac{11}{12}\epsilon(t)^{-1}-\frac{1}{6}\right],
\label{eq:(4.33)}\\
   \zeta_{13}^{(1)}(t)
   &=\frac{g_0^4}{(4\pi)^2}C_2(R)
   \left[-\frac{2}{3}\epsilon(t)^{-1}-\frac{7}{18}\right],
\label{eq:(4.34)}\\
   \zeta_{14}^{(1)}(t)
   &=\frac{g_0^4}{(4\pi)^2}C_2(R)
   \left[\frac{1}{3}\epsilon(t)^{-1}-\frac{7}{18}\right],
\label{eq:(4.35)}\\
   \zeta_{15}^{(1)}(t)
   &=\frac{g_0^4}{(4\pi)^2}C_2(R)
   \left[3\epsilon(t)^{-1}+\frac{5}{2}\right],
\label{eq:(4.36)}
\end{align}
where
\begin{equation}
   \epsilon(t)^{-1}\equiv\frac{1}{\epsilon}+\ln(8\pi t).
\label{eq:(4.37)}
\end{equation}
By considering the trace part of~$\Tilde{\mathcal{O}}_{1\mu\nu}$, from these we
further have
\begin{align}
   \zeta_{21}^{(1)}(t)
   &=0,
\label{eq:(4.38)}\\
   \zeta_{22}^{(1)}(t)
   &=\zeta_{11}^{(1)}(t)+\zeta_{12}^{(1)}(t)(4-2\epsilon)
   =\frac{g_0^2}{(4\pi)^2}C_2(G)\frac{7}{2},
\label{eq:(4.39)}\\
   \zeta_{23}^{(1)}(t)
   &=0,
\label{eq:(4.40)}\\
   \zeta_{24}^{(1)}(t)
   &=2\zeta_{13}^{(1)}(t)+\zeta_{14}^{(1)}(t)(4-2\epsilon)
   =\frac{g_0^4}{(4\pi)^2}C_2(R)(-3),
\label{eq:(4.41)}\\
   \zeta_{25}^{(1)}(t)
   &=\zeta_{15}^{(1)}(t)(4-2\epsilon)
   =\frac{g_0^4}{(4\pi)^2}C_2(R)
   \left[12\epsilon(t)^{-1}+4\right].
\label{eq:(4.42)}
\end{align}



\begin{figure}
\begin{minipage}{0.3\hsize}
\begin{center}
\includegraphics[width=3cm,clip]{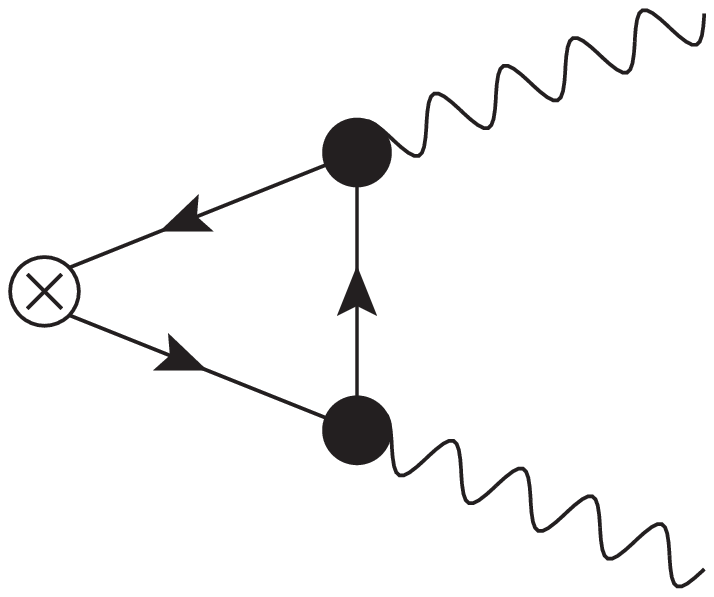}
\caption{B03}
\label{fig:31}
\end{center}
\end{minipage}
\begin{minipage}{0.3\hsize}
\begin{center}
\includegraphics[width=3cm,clip]{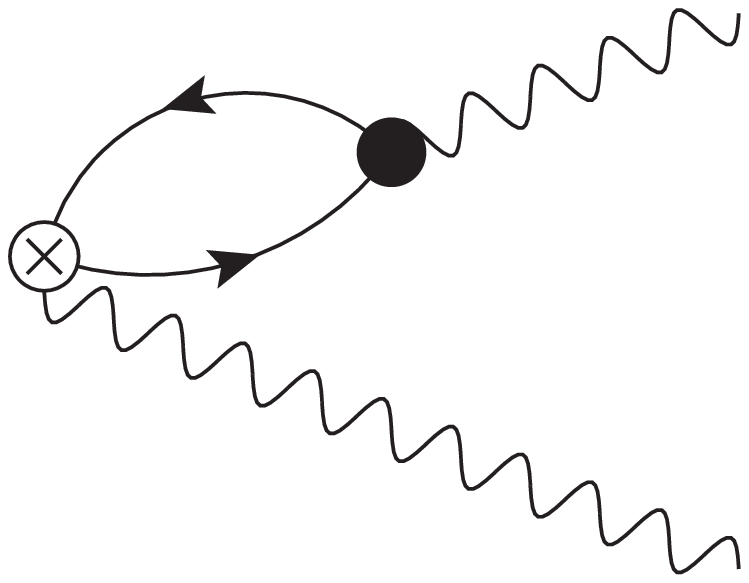}
\caption{B04}
\label{fig:32}
\end{center}
\end{minipage}
\begin{minipage}{0.3\hsize}
\begin{center}
\includegraphics[width=3cm,clip]{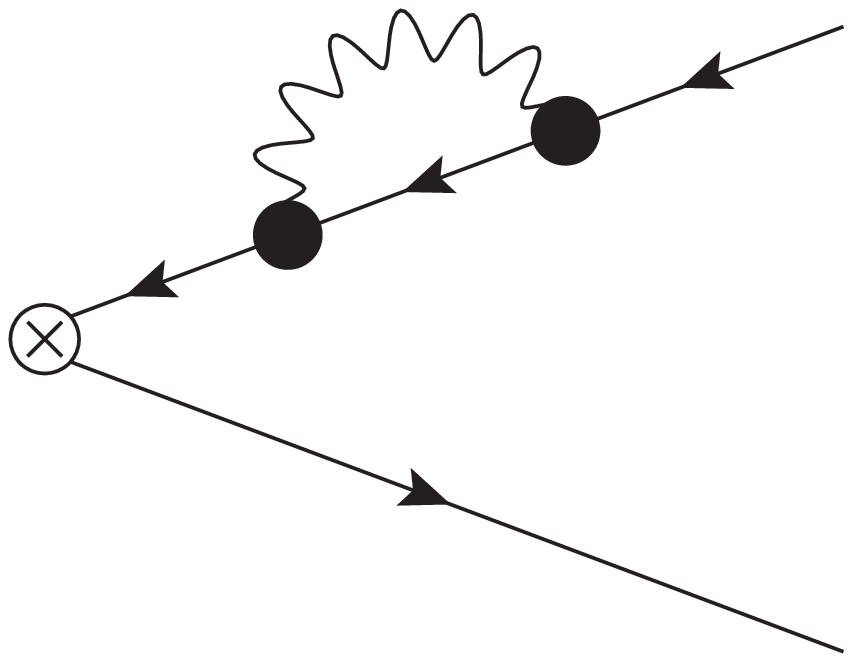}
\caption{B05}
\label{fig:33}
\end{center}
\end{minipage}
\end{figure}

\begin{figure}
\begin{minipage}{0.3\hsize}
\begin{center}
\includegraphics[width=3cm,clip]{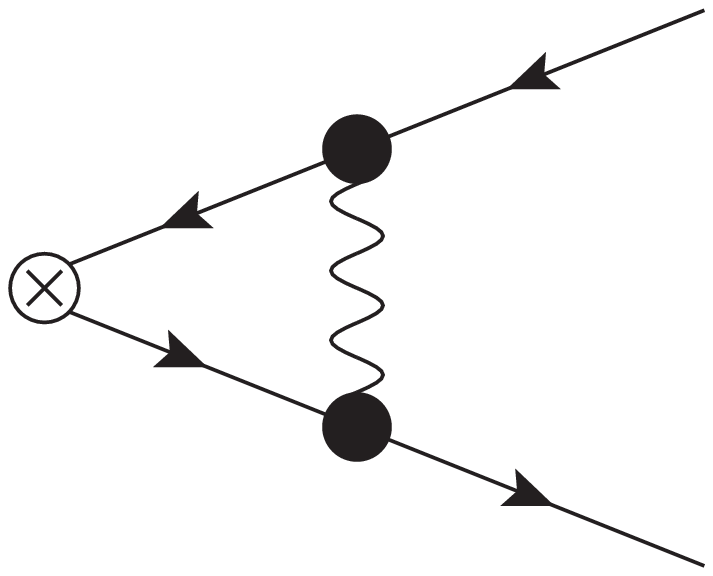}
\caption{B06}
\label{fig:34}
\end{center}
\end{minipage}
\begin{minipage}{0.3\hsize}
\begin{center}
\includegraphics[width=3cm,clip]{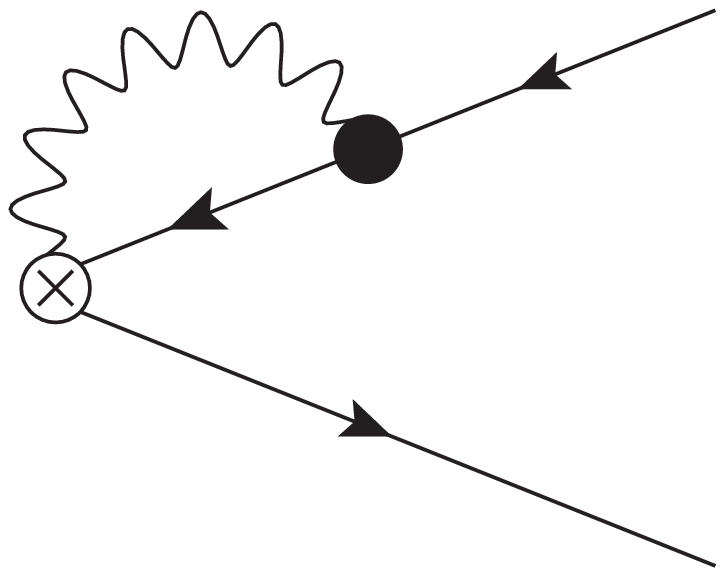}
\caption{B07}
\label{fig:35}
\end{center}
\end{minipage}
\begin{minipage}{0.3\hsize}
\begin{center}
\includegraphics[width=3cm,clip]{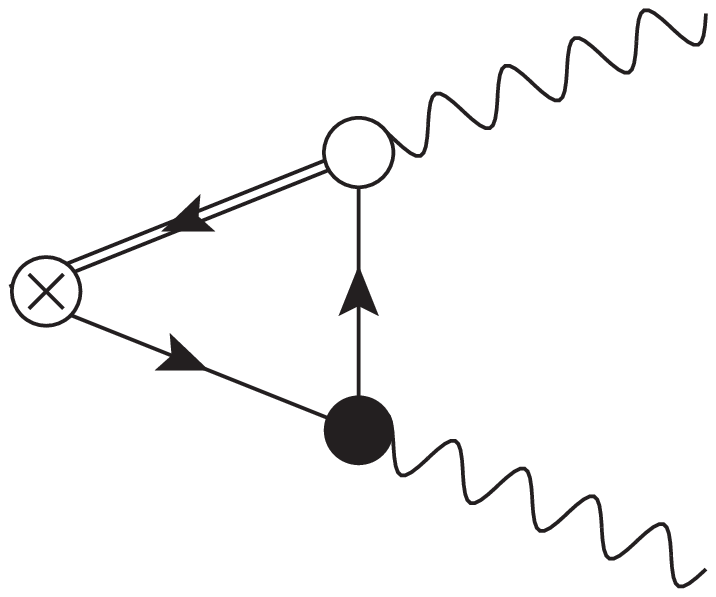}
\caption{B08}
\label{fig:36}
\end{center}
\end{minipage}
\end{figure}

\begin{figure}
\begin{minipage}{0.3\hsize}
\begin{center}
\includegraphics[width=3cm,clip]{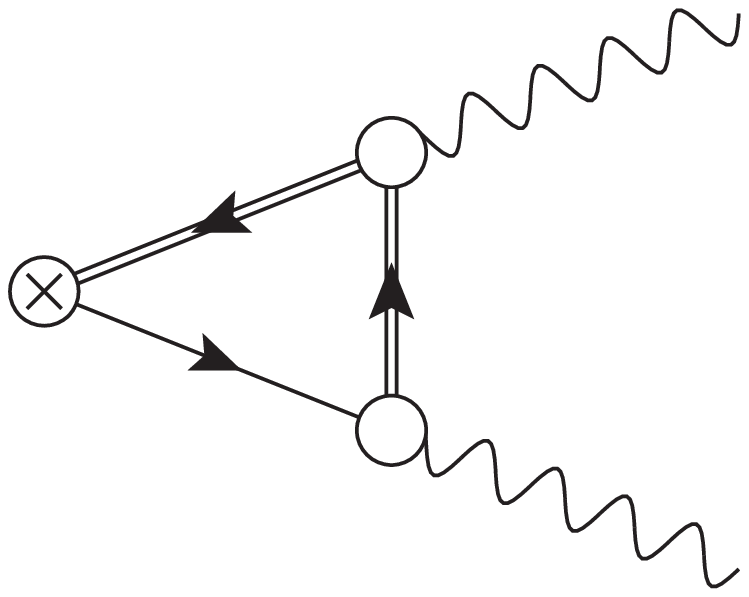}
\caption{B09}
\label{fig:37}
\end{center}
\end{minipage}
\begin{minipage}{0.3\hsize}
\begin{center}
\includegraphics[width=3cm,clip]{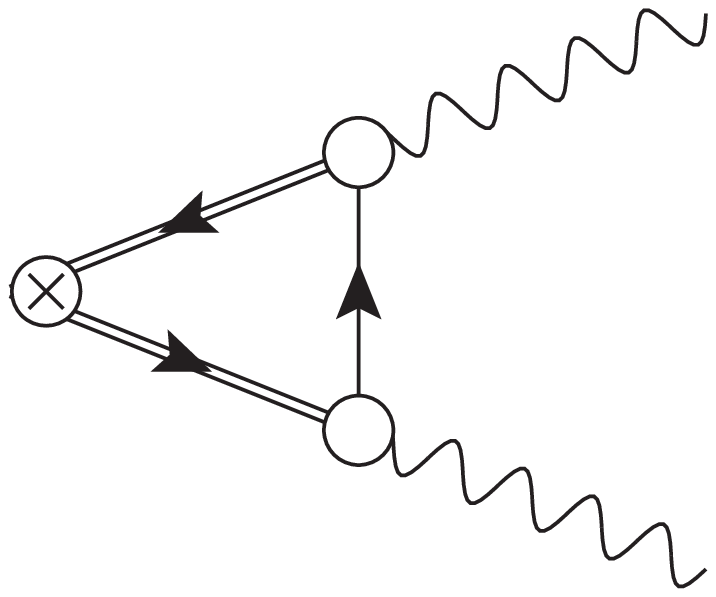}
\caption{B10}
\label{fig:38}
\end{center}
\end{minipage}
\begin{minipage}{0.3\hsize}
\begin{center}
\includegraphics[width=3cm,clip]{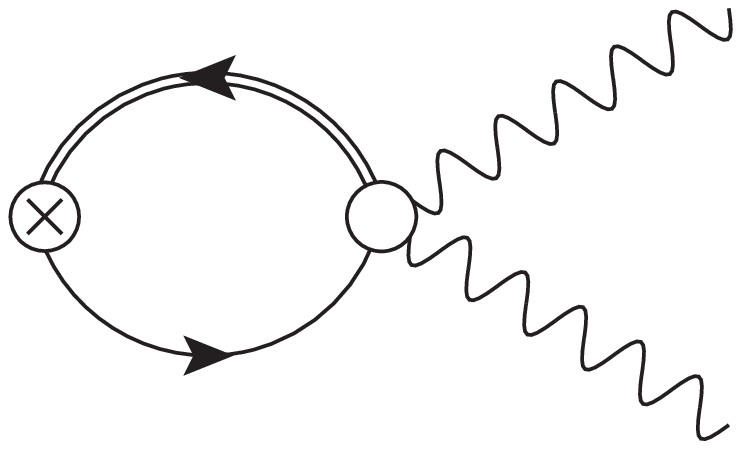}
\caption{B11}
\label{fig:39}
\end{center}
\end{minipage}
\end{figure}

\begin{figure}
\begin{minipage}{0.3\hsize}
\begin{center}
\includegraphics[width=3cm,clip]{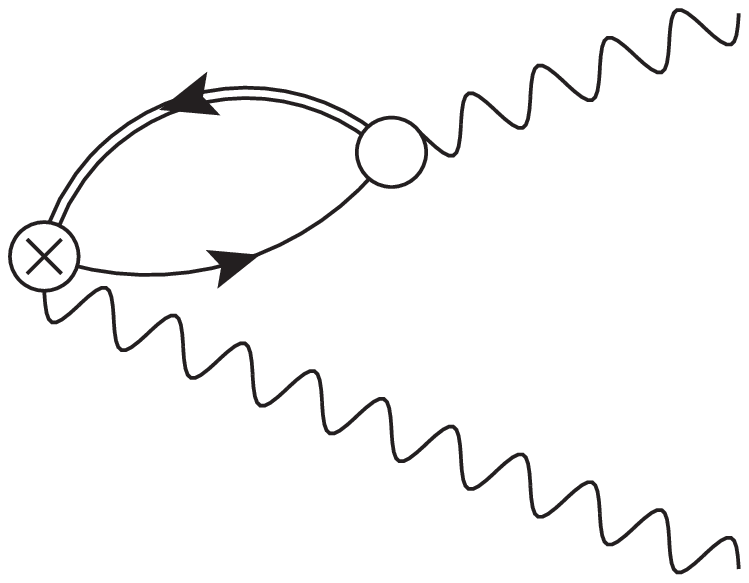}
\caption{B12}
\label{fig:40}
\end{center}
\end{minipage}
\begin{minipage}{0.3\hsize}
\begin{center}
\includegraphics[width=3cm,clip]{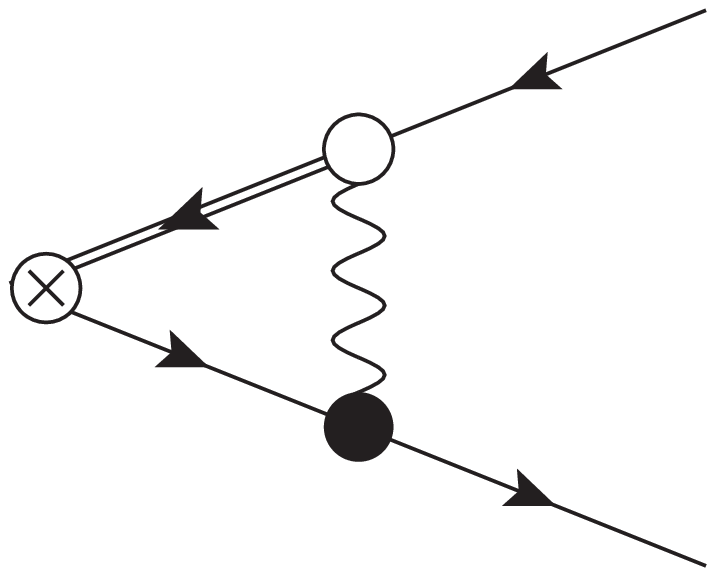}
\caption{B13}
\label{fig:41}
\end{center}
\end{minipage}
\begin{minipage}{0.3\hsize}
\begin{center}
\includegraphics[width=3cm,clip]{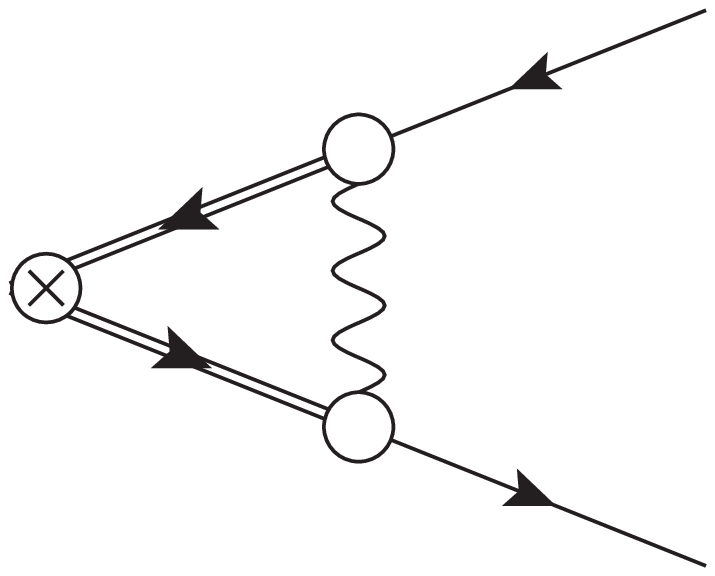}
\caption{B14}
\label{fig:42}
\end{center}
\end{minipage}
\end{figure}

\begin{figure}
\begin{minipage}{0.3\hsize}
\begin{center}
\includegraphics[width=3cm,clip]{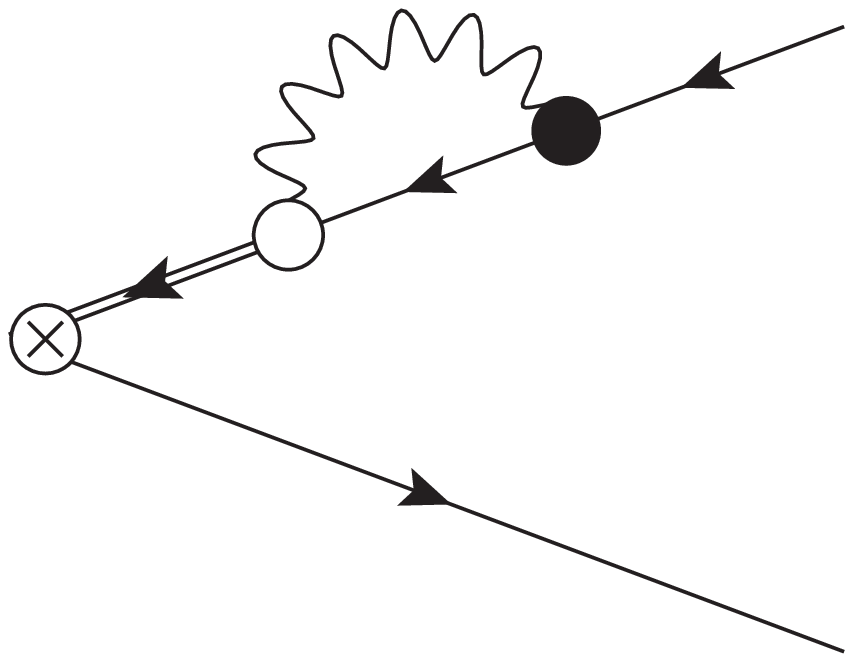}
\caption{B15}
\label{fig:43}
\end{center}
\end{minipage}
\begin{minipage}{0.3\hsize}
\begin{center}
\includegraphics[width=3cm,clip]{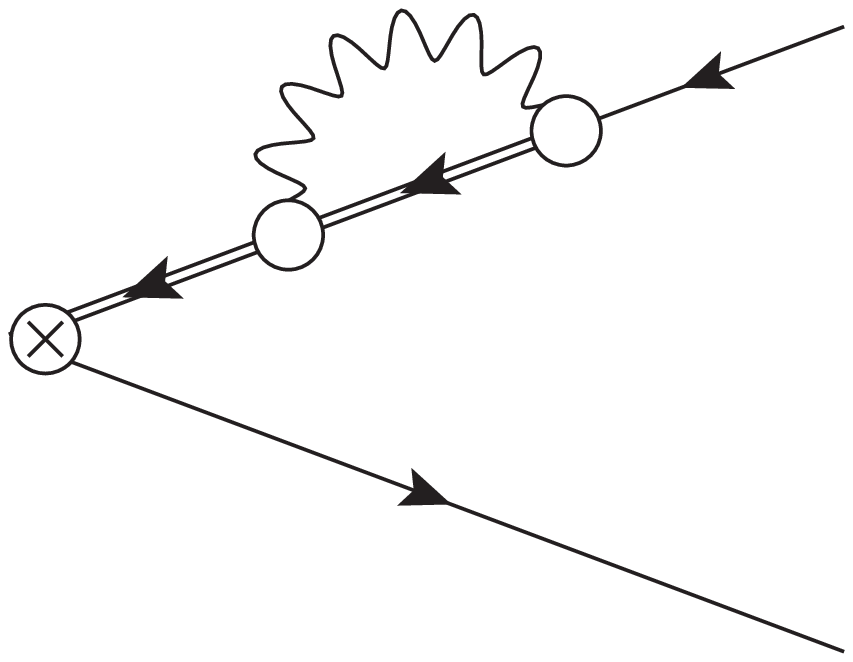}
\caption{B16}
\label{fig:44}
\end{center}
\end{minipage}
\begin{minipage}{0.3\hsize}
\begin{center}
\includegraphics[width=3cm,clip]{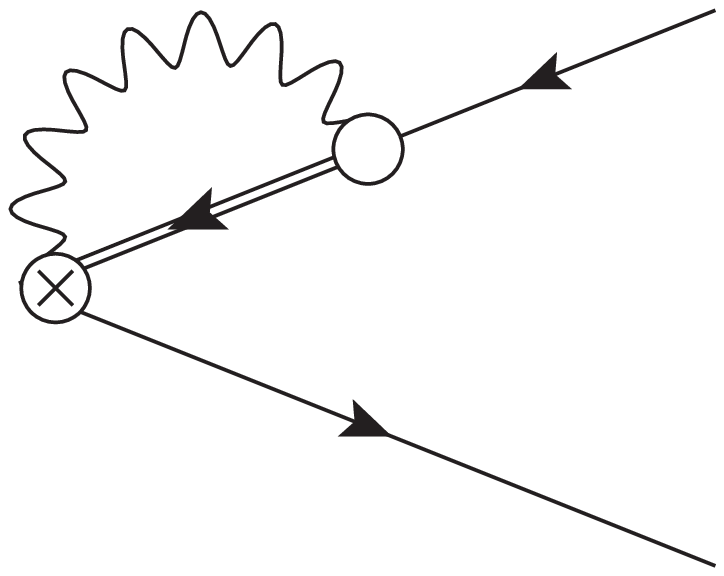}
\caption{B17}
\label{fig:45}
\end{center}
\end{minipage}
\end{figure}

\begin{figure}
\begin{minipage}{0.3\hsize}
\begin{center}
\includegraphics[width=3cm,clip]{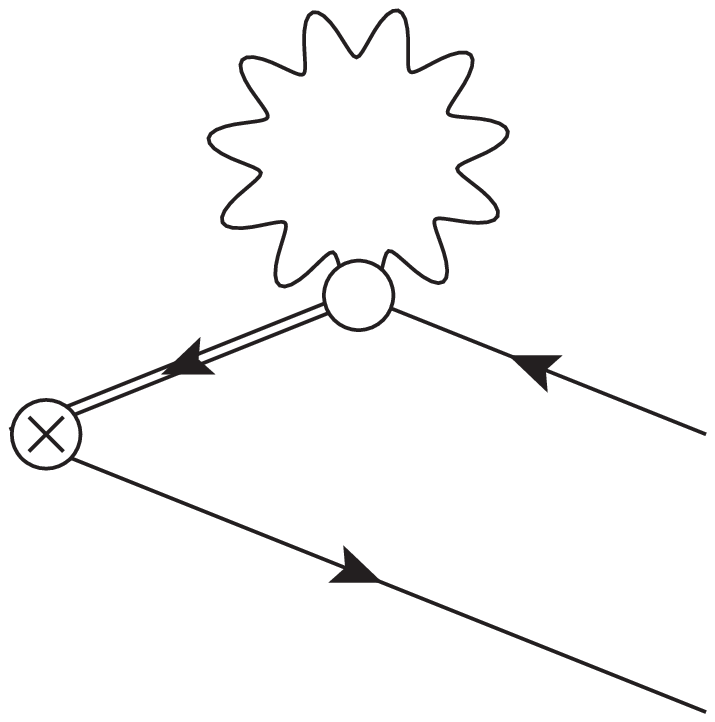}
\caption{B18}
\label{fig:46}
\end{center}
\end{minipage}
\end{figure}

For $\zeta_{3j}^{(1)}(t)$ with $j=1$ and~$2$, diagrams B03, B04, B08, B09, B10,
B11, and~B12 in~Figs.~\ref{fig:31}--\ref{fig:40} contribute. The contribution
of each diagram is tabulated in~Table~\ref{table:2}. For $\zeta_{3j}^{(1)}(t)$
with $j=3$, $4$, and~$5$, diagrams B06, B07, B13, B14, B15, B16, B17, and B18
in~Figs.~\ref{fig:34}--\ref{fig:46} contribute and their contributions are
tabulated in~Table~\ref{table:3}.
\begin{table}
\caption{$\zeta_{3j}^{(1)}$ in units of $\frac{1}{(4\pi)^2}T(R)N_{\mathrm{f}}$}
\label{table:2}
\begin{center}
\renewcommand{\arraystretch}{2.2}
\setlength{\tabcolsep}{20pt}
\begin{tabular}{crr}
\toprule
 diagram & \multicolumn{1}{c}{$\zeta_{31}^{(1)}(t)$}
 & \multicolumn{1}{c}{$\zeta_{32}^{(1)}(t)$} \\
\midrule
B03  & $-\dfrac{16}{3}\epsilon(t)^{-1}-\dfrac{64}{9}$ & $\dfrac{4}{3}\epsilon(t)^{-1}+\dfrac{25}{9}$ \\
B04  & $0$ & $0$ \\
B08  & $0$ & $-\dfrac{1}{3}$ \\
B09  & $-\dfrac{5}{12}$ & $-\dfrac{23}{144}$ \\
B10  & $\dfrac{5}{12}$ & $-\dfrac{1}{16}$ \\
B11  & $\dfrac{10}{9}$ & $\dfrac{7}{9}$ \\
B12  & $0$ & $0$ \\
\bottomrule
\end{tabular}
\end{center}
\end{table}
\begin{table}
\caption{$\zeta_{3j}^{(1)}$ in units of $\frac{g_0^2}{(4\pi)^2}C_2(R)$}
\label{table:3}
\begin{center}
\renewcommand{\arraystretch}{2.2}
\setlength{\tabcolsep}{20pt}
\begin{tabular}{crrr}
\toprule
 diagram & \multicolumn{1}{c}{$\zeta_{33}^{(1)}(t)$}
 & \multicolumn{1}{c}{$\zeta_{34}^{(1)}(t)$}
 & \multicolumn{1}{c}{$\zeta_{35}^{(1)}(t)$} \\
\midrule
 B06 & $-\dfrac{1}{3}\epsilon(t)^{-1}+\dfrac{1}{18}$ & $\dfrac{2}{3}\epsilon(t)^{-1}+\dfrac{5}{9}$ & $8\epsilon(t)^{-1}+8$ \\
 B07 & $2\epsilon(t)^{-1}+2$ & $-2\epsilon(t)^{-1}-2$ & $-8\epsilon(t)^{-1}-8$ \\
 B13 & $0$ & $1$ & $0$ \\
 B14 & $0$ & $0$ & $0$ \\
 B15 & $2\epsilon(t)^{-1}+2$ & $0$ & $0$ \\
 B16 & $0$ & $0$ & $0$ \\
 B17 & $-2$ & $0$ & $0$ \\
 B18 & $-4\epsilon(t)^{-1}-2$ & $0$ & $0$ \\
\bottomrule
\end{tabular}
\end{center}
\end{table}
As the sum of these contributions, we have
\begin{align}
   \zeta_{31}^{(1)}(t)
   &=\frac{1}{(4\pi)^2}T(R)N_{\mathrm{f}}
   \left[-\frac{16}{3}\epsilon(t)^{-1}-6\right],
\label{eq:(4.43)}\\
   \zeta_{32}^{(1)}(t)
   &=\frac{1}{(4\pi)^2}T(R)N_{\mathrm{f}}
   \left[\frac{4}{3}\epsilon(t)^{-1}+3\right],
\label{eq:(4.44)}\\
   \zeta_{33}^{(1)}(t)
   &=\frac{g^2}{(4\pi)^2}C_2(R)
   \left[3\frac{1}{\epsilon}-\Phi(t)\right]
   +\frac{g_0^2}{(4\pi)^2}C_2(R)
   \left[-\frac{1}{3}\epsilon(t)^{-1}+\frac{1}{18}\right],
\label{eq:(4.45)}\\
   \zeta_{34}^{(1)}(t)
   &=\frac{g_0^2}{(4\pi)^2}C_2(R)
   \left[-\frac{4}{3}\epsilon(t)^{-1}-\frac{4}{9}\right],
\label{eq:(4.46)}\\
   \zeta_{35}^{(1)}(t)
   &=0,
\label{eq:(4.47)}
\end{align}
where in $\zeta_{33}^{(1)}(t)$~\eqref{eq:(4.45)} the first term in the
right-hand side comes from the conversion from the un-ringed fields to the
ringed fields in~Eqs.~\eqref{eq:(3.20)} and~\eqref{eq:(3.21)}---recall
Eq.~\eqref{eq:(3.25)}; the combination~$\Phi(t)$ is given
by~Eq.~\eqref{eq:(3.27)}. From these, we further have
\begin{align}
   \zeta_{41}^{(1)}(t)
   &=0,
\label{eq:(4.48)}\\
   \zeta_{42}^{(1)}(t)
   &=\frac{1}{2}\zeta_{31}^{(1)}(t)+\frac{1}{2}\zeta_{32}^{(1)}(t)(4-2\epsilon)
   =\frac{1}{(4\pi)^2}T(R)N_{\mathrm{f}}
   \frac{5}{3},
\label{eq:(4.49)}\\
   \zeta_{43}^{(1)}(t)
   &=0,
\label{eq:(4.50)}\\
   \zeta_{44}^{(1)}(t)
   &=\zeta_{33}^{(1)}(t)+\frac{1}{2}\zeta_{34}^{(1)}(t)(4-2\epsilon)
\notag\\
   &=\frac{g^2}{(4\pi)^2}C_2(R)
   \left[3\frac{1}{\epsilon}-\Phi(t)\right]
   +\frac{g_0^2}{(4\pi)^2}C_2(R)
   \left[-3\epsilon(t)^{-1}+\frac{1}{2}\right],
\label{eq:(4.51)}\\
   \zeta_{45}^{(1)}(t)
   &=0.
\label{eq:(4.52)}
\end{align}

Finally,
\begin{equation}
   \zeta_{51}^{(1)}(t)
   =\zeta_{52}^{(1)}(t)=\zeta_{53}^{(1)}(t)=\zeta_{54}^{(1)}(t)=0,
\label{eq:(4.53)}
\end{equation}
and $\zeta_{55}^{(1)}(t)$ is given by the sum of the contributions of one-loop
diagrams in~Table~\ref{table:4} and the conversion factor to the ringed fields:
\begin{table}
\caption{$\zeta_{55}^{(1)}(t)$ in units of~$\frac{g_0^2}{(4\pi)^2}C_2(R)$.}
\label{table:4}
\begin{center}
\renewcommand{\arraystretch}{2.2}
\setlength{\tabcolsep}{20pt}
\begin{tabular}{cr}
\toprule
 diagram & \multicolumn{1}{c}{$\zeta_{55}^{(1)}(t)$} \\
\midrule
 B06 & $-4\epsilon(t)^{-1}-2$ \\
 B13 & $0$ \\
 B14 & $0$ \\
 B15 & $2\epsilon(t)^{-1}+2$ \\
 B16 & $0$ \\
 B18 & $-4\epsilon(t)^{-1}-2$ \\
\bottomrule
\end{tabular}
\end{center}
\end{table}
\begin{align}
   \zeta_{55}^{(1)}(t)
   =\frac{g^2}{(4\pi)^2}C_2(R)
   \left[6\frac{1}{\epsilon}-\Phi(t)\right]
   +\frac{g_0^2}{(4\pi)^2}C_2(R)
   \left[-6\epsilon(t)^{-1}-2\right].
\label{eq:(4.54)}
\end{align}

We have now obtained all $\zeta_{ij}^{(1)}(t)$ in~Eqs.~\eqref{eq:(4.24)}
and~\eqref{eq:(4.25)}. Then, since the matrix~$\zeta_{ij}(t)$ in the tree-level
approximation is a unit matrix, it is straightforward to invert the
matrix~$\zeta_{ij}(t)$ in the one-loop approximation; Eqs.~\eqref{eq:(4.15)}
and~\eqref{eq:(4.16)} thus yield
\begin{align}
   c_1(t)&=\frac{1}{g_0^2}
   \left[1-\zeta_{11}^{(1)}(t)\right]
   -\frac{1}{4}\zeta_{31}^{(1)}(t),
\label{eq:(4.55)}\\
   c_2(t)&=\frac{1}{g_0^2}
   \left(-\frac{1}{2}\epsilon\right)\zeta_{12}^{(1)}(t)
   +\left(\frac{3}{4}-\frac{1}{2}\epsilon\right)\zeta_{32}^{(1)}(t)
   +\frac{3}{16}\zeta_{31}^{(1)}(t),
\label{eq:(4.56)}\\
   c_3(t)&=\left\{\frac{1}{g_0^2}
   \left[-\zeta_{13}^{(1)}(t)\right]
   +\frac{1}{4}-\frac{1}{4}\zeta_{33}^{(1)}(t)\right\}Z(\epsilon)^{-1},
\label{eq:(4.57)}\\
   c_4(t)&=\left\{\frac{1}{g_0^2}
   \left[-\frac{1}{2}\epsilon\zeta_{14}^{(1)}(t)
   -\frac{3}{2}\zeta_{13}^{(1)}(t)\right]
   +\left(\frac{3}{4}-\frac{1}{2}\epsilon\right)\zeta_{34}^{(1)}(t)\right\}
   Z(\epsilon)^{-1},
\label{eq:(4.58)}\\
   c_5(t)&=\left\{\frac{1}{g_0^2}
   \left[-\frac{1}{2}\epsilon\zeta_{15}^{(1)}(t)\right]
   -1+\zeta_{55}^{(1)}(t)\right\}Z(\epsilon)^{-1}.
\label{eq:(4.59)}
\end{align}
Then, by using the renormalized gauge coupling in the $\text{MS}$
scheme~\eqref{eq:(A1)}, to the one-loop order, we have (for~$\epsilon\to0$)
\begin{align}
   c_1(t)&
   =\frac{1}{g^2}-b_0\ln(8\pi\mu^2t)-\frac{1}{(4\pi)^2}
   \left[\frac{7}{3}C_2(G)
   -\frac{3}{2}T(R)N_{\mathrm{f}}\right],
\label{eq:(4.60)}\\
   c_2(t)&
   =\frac{1}{8}
   \frac{1}{(4\pi)^2}
   \left[\frac{11}{3}C_2(G)
   +\frac{11}{3}T(R)N_{\mathrm{f}}\right],
\label{eq:(4.61)}\\
   c_3(t)&
   =\frac{1}{4}\left\{1+\frac{g^2}{(4\pi)^2}C_2(R)
   \left[\frac{3}{2}+\ln(432)\right]\right\},
\label{eq:(4.62)}\\
   c_4(t)&=\frac{1}{8}d_0g^2,
\label{eq:(4.63)}\\
   c_5(t)&=-\left\{1+\frac{g^2}{(4\pi)^2}C_2(R)
   \left[3\ln(8\pi\mu^2t)+\frac{7}{2}+\ln(432)\right]\right\}.
\label{eq:(4.64)}
\end{align}
Since the $c_i(t)$ in~Eq.~\eqref{eq:(4.14)} connect the finite energy--momentum
tensor and UV-finite local products $\Tilde{\mathcal{O}}_{i\mu\nu}(t,x)$
constructed from (ringed) flowed fields, they should be UV finite. That our
explicit one-loop calculation of~$c_i(t)$ confirms this finiteness is quite
reassuring.

If one prefers the $\overline{\text{MS}}$ scheme instead of the $\text{MS}$
scheme assumed in above expressions, it suffices to make the replacement
\begin{equation}
    \mu^2\to\frac{\mathrm{e}^{\gamma_{\mathrm{E}}}}{4\pi}\mu^2,
\label{eq:(4.65)}
\end{equation}
where $\gamma_{\mathrm{E}}$ is Euler's constant.

\subsection{A consistency check: The trace anomaly}
It is interesting to see that Eq.~\eqref{eq:(4.14)} with $c_i(t)$
in~Eqs.~\eqref{eq:(4.60)}--\eqref{eq:(4.64)} in fact reproduces the trace
anomaly~\eqref{eq:(2.11)} in the one-loop approximation. At first brief glance,
$c_2(t)$ in~Eq.~\eqref{eq:(4.61)} is incompatible with the correct trace
anomaly, because $\delta_{\mu\nu}[\Tilde{\mathcal{O}}_{1\mu\nu}
-(1/4)\Tilde{\mathcal{O}}_{2\mu\nu}]=0$ for~$D=4$ and
$\delta_{\mu\nu}\Tilde{\mathcal{O}}_{2\mu\nu}
=4\left\{F_{\rho\sigma}^aF_{\rho\sigma}^a\right\}_R(x)+O(g^2)$ [the last equality
follows from Eq.~\eqref{eq:(A17)}]. On the other hand, $4c_2(t)$
from~Eq.~\eqref{eq:(4.61)} is not identical to~$b_0/2$, where $b_0$ is the
first coefficient of the beta function~\eqref{eq:(2.16)}, the correct one-loop
coefficient of the trace anomaly.

This is a premature judgment, however. In fact, $\zeta_{42}^{(1)}(t)$
in~Eq.~\eqref{eq:(4.49)} shows that there exists an operator mixing of the form
\begin{equation}
   \mathring{\Bar{\chi}}(t,x)
   \overleftrightarrow{\Slash{D}}
   \mathring{\chi}(t,x)
   =\left\{\Bar{\psi}\overleftrightarrow{\Slash{D}}\psi\right\}_R(x)
   +\frac{1}{(4\pi)^2}T(R)N_{\mathrm{f}}\frac{5}{3}
   \left\{F_{\rho\sigma}^aF_{\rho\sigma}^a\right\}_R(x)
   +\dotsb.
\label{eq:(4.66)}
\end{equation}
Then the last term precisely fills the difference between $4c_2(t)$
and~$b_0/2$.

In this way, to the one-loop order, we have
\begin{align}
   \delta_{\mu\nu}\left\{T_{\mu\nu}\right\}_R(x)
   &=\frac{1}{2}\frac{1}{(4\pi)^2}
   \left[\frac{11}{3}C_2(G)-\frac{4}{3}T(R)N_{\mathrm{f}}\right]
   \left\{F_{\rho\sigma}^aF_{\rho\sigma}^a\right\}_R(x)
\notag\\
   &\qquad{}
   -\frac{3}{2}
   \left\{\Bar{\psi}\overleftrightarrow{\Slash{D}}\psi\right\}_R(x)
   -\left[4+\frac{g^2}{(4\pi)^2}6C_2(R)\right]
   m\left\{\Bar{\psi}\psi\right\}_R(x).
\label{eq:(4.67)}
\end{align}
This reproduces the trace anomaly~\eqref{eq:(2.11)} in the one-loop level if
one uses the equation of motion of renormalized field,
\begin{equation}
   \left\{\Bar{\psi}\overleftrightarrow{\Slash{D}}\psi\right\}_R(x)
   =-2m\left\{\Bar{\psi}\psi\right\}_R(x),
\label{eq:(4.68)}
\end{equation}
whose use is justified when there is no other operator in the point~$x$ as we
are assuming. We observe that our one-loop result
in~Eqs.~\eqref{eq:(4.60)}--\eqref{eq:(4.64)} is consistent with the trace
anomaly.

In Ref.~\cite{Suzuki:2013gza}, for the pure Yang--Mills theory, the
next-to-leading (two-loop order) term in~$c_2(t)$ was determined as
\begin{equation}
   c_2(t)=\frac{1}{8}b_0
   -\frac{1}{8}b_0
   \left[\frac{1}{(4\pi)^2}G_2(G)\frac{7}{2}-\frac{b_1}{b_0}\right]g^2,
\label{eq:(4.69)}
\end{equation}
where $b_0=[1/(4\pi)^2](11/3)C_2(G)$ and~$b_1=[1/(4\pi)^4](34/3)C_2(G)^2$, by
imposing that the expression~\eqref{eq:(4.14)} reproduces the trace
anomaly~\eqref{eq:(2.11)} to the two-loop order. For the present system with
fermions, however, it seems that this requirement alone cannot fix the
next-to-leading terms in~$c_2(t)$ and in~$c_4(t)$; so we are content with the
one-loop formulas, Eqs.~\eqref{eq:(4.60)}--\eqref{eq:(4.64)}, in the present
paper treating a system containing fermions.

\subsection{Master formula}
From~Eq.~\eqref{eq:(4.14)}, the energy--momentum tensor is given by the $t\to0$
limit,
\begin{align}
   \left\{T_{\mu\nu}\right\}_R(x)
   &=\lim_{t\to0}\biggl\{c_1(t)\left[
   \Tilde{\mathcal{O}}_{1\mu\nu}(t,x)
   -\frac{1}{4}\Tilde{\mathcal{O}}_{2\mu\nu}(t,x)
   \right]
\notag\\
   &\qquad\qquad{}
   +c_2(t)\left[
   \Tilde{\mathcal{O}}_{2\mu\nu}(t,x)
   -\left\langle\Tilde{\mathcal{O}}_{2\mu\nu}(t,x)\right\rangle
   \right]
\notag\\
   &\qquad\qquad\qquad{}
   +c_3(t)\left[
   \Tilde{\mathcal{O}}_{3\mu\nu}(t,x)
   -2\Tilde{\mathcal{O}}_{4\mu\nu}(t,x)
   -\left\langle
   \Tilde{\mathcal{O}}_{3\mu\nu}(t,x)
   -2\Tilde{\mathcal{O}}_{4\mu\nu}(t,x)
   \right\rangle
   \right]
\notag\\
   &\qquad\qquad\qquad\qquad{}
   +c_4(t)\left[
   \Tilde{\mathcal{O}}_{4\mu\nu}(t,x)
   -\left\langle\Tilde{\mathcal{O}}_{4\mu\nu}(t,x)\right\rangle
   \right]
\notag\\
   &\qquad\qquad\qquad\qquad\qquad{}
   +c_5(t)\left[
   \Tilde{\mathcal{O}}_{5\mu\nu}(t,x)
   -\left\langle\Tilde{\mathcal{O}}_{5\mu\nu}(t,x)\right\rangle
   \right]\biggr\},
\label{eq:(4.70)}
\end{align}
where operators in the right-hand side are given
by~Eqs.~\eqref{eq:(4.1)}--\eqref{eq:(4.5)}. One may further use the identities
\begin{equation}
   2\left\langle\Tilde{\mathcal{O}}_{3\mu\nu}(t,x)\right\rangle
   =\left\langle\Tilde{\mathcal{O}}_{4\mu\nu}(t,x)\right\rangle
   =\frac{-2\dim(R)N_{\mathrm{f}}}{(4\pi)^2t^2}\delta_{\mu\nu}
\label{eq:(4.71)}
\end{equation}
to make the expression a little simpler. Applying the consequence of the
renormalization group argument, Eqs.~\eqref{eq:(4.22)}--\eqref{eq:(4.23)},
to~Eqs.~\eqref{eq:(4.60)}--\eqref{eq:(4.64)}, we have
\begin{align}
   c_1(t)&
   =\frac{1}{\Bar{g}(1/\sqrt{8t})^2}
   -b_0\ln\pi-\frac{1}{(4\pi)^2}
   \left[\frac{7}{3}C_2(G)
   -\frac{3}{2}T(R)N_{\mathrm{f}}\right],
\label{eq:(4.72)}\\
   c_2(t)&
   =\frac{1}{8}
   \frac{1}{(4\pi)^2}
   \left[\frac{11}{3}C_2(G)
   +\frac{11}{3}T(R)N_{\mathrm{f}}\right],
\label{eq:(4.73)}\\
   c_3(t)&
   =\frac{1}{4}\left\{1+\frac{\Bar{g}(1/\sqrt{8t})^2}{(4\pi)^2}C_2(R)
   \left[\frac{3}{2}+\ln(432)\right]\right\},
\label{eq:(4.74)}\\
   c_4(t)&=\frac{1}{8}d_0\Bar{g}(1/\sqrt{8t})^2,
\label{eq:(4.75)}\\
   c_5(t)&=-\frac{\Bar{m}(1/\sqrt{8t})}{m}
   \left\{1+\frac{\Bar{g}(1/\sqrt{8t})^2}{(4\pi)^2}C_2(R)
   \left[3\ln\pi+\frac{7}{2}+\ln(432)\right]\right\},
\label{eq:(4.76)}
\end{align}
where $\Bar{g}(q)$ is the running gauge coupling in the $\text{MS}$ scheme.
For going from the $\text{MS}$ scheme to the $\overline{\text{MS}}$ scheme, it
suffices to make the following replacement corresponding
to~Eq.~\eqref{eq:(4.65)}, 
\begin{equation}
   \ln\pi\to\gamma_{\mathrm{E}}-2\ln2,
\label{eq:(4.77)}
\end{equation}
in the above expressions. Equation~\eqref{eq:(4.70)} with
Eqs.~\eqref{eq:(4.72)}--\eqref{eq:(4.76)} is our main result. Note that the
renormalized mass parameter~$m$ in~$c_5(t)$~\eqref{eq:(4.76)} and that
in~$\Tilde{\mathcal{O}}_{5\mu\nu}(t,x)$~\eqref{eq:(4.5)} are redundant
in~Eq.~\eqref{eq:(4.70)} because they are cancelled out in the product. For the
running mass parameter~$\Bar{m}(1/\sqrt{8t})$ in~$c_5(t)$ for~$t\to0$, one may
use the relation
\begin{align}
   \Bar{m}(q)&=[2b_0\Bar{g}(q)^2]^{d_0/2b_0}
   \exp\left\{\int_0^{\Bar{g}(q)}\mathrm{d}g\,
   \left[-\frac{\gamma_m(g)}{\beta(g)}-\frac{d_0}{b_0g}\right]\right\}m^\infty
\notag\\
   &=\left(\frac{2}{\ell}\right)^{d_0/2b_0}
   \left[1-\frac{d_0b_1}{2b_0^3\ell}(1+\ln\ell)+\frac{d_1}{2b_0^2\ell}
   +O(\ell^{-2})\right]m^\infty,\qquad\ell\equiv\ln(q^2/\Lambda^2),
\label{eq:(4.78)}
\end{align}
where $m^\infty$ denotes the renormalization group invariant mass. For the
massive fermion, $m^\infty$~may be determined by using the method established
in~Ref.~\cite{Capitani:1998mq}, for example. For the massless fermion,
$m^\infty=0$ and we can simply discard the last line of~Eq.~\eqref{eq:(4.70)}.

\section{Equation of motion in the small flow-time limit}
\label{sec:5}
Let us consider the following representations for small flow-time:
\begin{align}
   &\delta_{\mu\nu}\left[\Bar{\psi}(x)\overleftrightarrow{\Slash{D}}\psi(x)
   +2m_0\Bar{\psi}(x)\psi(x)\right]
\notag\\
   &=d_2(t)\Tilde{\mathcal{O}}_{2\mu\nu}(t,x)
   +d_4(t)\Tilde{\mathcal{O}}_{4\mu\nu}(t,x)
   +d_5(t)\Tilde{\mathcal{O}}_{5\mu\nu}+O(t),
\label{eq:(5.1)}
\end{align}
and
\begin{equation}
   \delta_{\mu\nu}m_0\Bar{\psi}(x)\psi(x)
   =e_5(t)\Tilde{\mathcal{O}}_{5\mu\nu}(t,x)+O(t),
\label{eq:(5.2)}
\end{equation}
where it is understood that the vacuum expectation values are subtracted on
both sides of the equations. By a renormalization group argument identical to
that which led to~Eqs.~\eqref{eq:(4.22)} and~\eqref{eq:(4.23)} and the one-loop
calculation in~Sect.~\ref{sec:4}, for~$t\to0$ we have
\begin{align}
   d_2(t)&=-\frac{1}{(4\pi)^2}T(R)N_{\mathrm{f}}\frac{5}{3},
\label{eq:(5.3)}\\
   d_4(t)&=1+\frac{\Bar{g}(1/\sqrt{8t})^2}{(4\pi)^2}C_2(R)
   \left[-\frac{1}{2}+\ln(432)\right],
\label{eq:(5.4)}\\
   d_5(t)&=\frac{\Bar{m}(1/\sqrt{8t})}{m}
   2\left\{1+\frac{\Bar{g}(1/\sqrt{8t})^2}{(4\pi)^2}C_2(R)
   \left[3\ln\pi+2+\ln(432)\right]\right\},
\label{eq:(5.5)}
\end{align}
and $e_5(t)=(1/2)d_5(t)$. We note that since the left-hand side
of~Eq.~\eqref{eq:(5.1)} is proportional to the equation of motion of the
fermion field, when the position~$x$ does not coincide with positions of other
operators in the position space, we may set the combination to zero (the
Schwinger--Dyson equation). This implies that we can make the replacement (the
subtraction of the vacuum expectation value is understood)
\begin{align}
   \Tilde{\mathcal{O}}_{4\mu\nu}(t,x)
   &\to\frac{1}{(4\pi)^2}T(R)N_{\mathrm{f}}\frac{5}{3}
   \Tilde{\mathcal{O}}_{2\mu\nu}(t,x)
\notag\\
   &\qquad{}
   -\frac{\Bar{m}(1/\sqrt{8t})}{m}2
   \left[1+\frac{\Bar{g}(1/\sqrt{8t})^2}{(4\pi)^2}C_2(R)
   \left(3\ln\pi+\frac{5}{2}\right)\right]\Tilde{\mathcal{O}}_{5\mu\nu}(t,x)
\label{eq:(5.6)}
\end{align}
in the master formula~\eqref{eq:(4.70)} [for the $\overline{\text{MS}}$ scheme,
one makes the substitution~\eqref{eq:(4.77)}], because throughout this paper we
are assuming that the energy--momentum tensor~$\{T_{\mu\nu}\}_R(x)$ is separated
from other operators in correlation functions. This makes the expression of the
energy--momentum tensor somewhat simpler.

If one is interested in the trace part of the energy--momentum tensor, that is,
the total divergence of the dilatation current, the above procedure leads to
\begin{align}
   \delta_{\mu\nu}\left\{T_{\mu\nu}\right\}_R(x)
   &=\lim_{t\to0}
   \biggl(\frac{1}{2}b_0
   \left[
   G_{\rho\sigma}^a(t,x)G_{\rho\sigma}^a(t,x)
   -\left\langle G_{\rho\sigma}^a(t,x)G_{\rho\sigma}^a(t,x)\right\rangle
   \right]
\notag\\
   &\qquad\qquad{}
   -\left\{1+\frac{\Bar{g}(1/\sqrt{8t})^2}{(4\pi)^2}C_2(R)
   \left[3\ln\pi+8+\ln(432)\right]\right\}
\notag\\
   &\qquad\qquad\qquad\qquad{}
   \times\Bar{m}(1/\sqrt{8t})\left[
   \mathring{\Bar{\chi}}(t,x)\mathring{\chi}(t,x)
   -\left\langle
   \mathring{\Bar{\chi}}(t,x)\mathring{\chi}(t,x)\right\rangle
   \right]
   \biggr),
\label{eq:(5.7)}
\end{align}
which is quite analogous to the trace anomaly~\eqref{eq:(2.11)}. For the
massless fermion, $\Bar{m}(1/\sqrt{8t})=0$ and we end up with a quite simple
expression for the trace part of the energy--momentum tensor.

\section{Conclusion}
\label{sec:6}
In the present paper, on the basis of the Yang--Mills gradient flow, we
constructed a formula~\eqref{eq:(4.70)} that provides a possible method to
compute correlation functions containing the energy--momentum tensor in lattice
gauge theory with fermions. This is a natural generalization of the
construction in~Ref.~\cite{Suzuki:2013gza} for the pure Yang--Mills theory.
Although the feasibility of the application in lattice Monte Carlo simulations
remains to be carefully investigated, the experience in the thermodynamics of
the quenched QCD~\cite{Asakawa:2013laa} strongly indicates that, even with
presently available lattice parameters, there exists a window~\eqref{eq:(1.2)}
within which one can reliably carry out the extrapolation for~$t\to0$
in~Eq.~\eqref{eq:(4.70)}. We expect various applications of the present
formulation. One is the application in many-flavor gauge theories with an
infrared fixed point (which are subject to recent active investigations; see
contributions in the last lattice conference~\cite{Itou:2013faa,Fodor:2014pqa}
for recent reviews).

\section*{Acknowledgments}
We would like to thank following people for valuable remarks:
Sinya Aoki,
Masayuki Asakawa,
Michael G. Endres,
Kazuo Fujikawa,
Leonardo Giusti,
Shoji Hashimoto,
Tetsuo Hatsuda,
Etsuko Itou,
Yoshio Kikukawa,
Masakiyo Kitazawa,
Tetsuya Onogi,
Giancarlo Rossi,
Shoichi Sasaki,
Yusuke Taniguchi,
and especially
Martin L\"uscher
for also making his private research notes available.
The work of H.~S. is supported in part by Grant-in-Aid for Scientific
Research~23540330.

\appendix

\section{One-loop renormalization in the MS scheme}
\label{sec:A}
\subsection{Parameters, elementary fields}
The gauge coupling:
\begin{equation}
   g_0^2=\mu^{2\epsilon}g^2
   \left\{
   1+\frac{g^2}{(4\pi)^2}\left[-\frac{11}{3}C_2(G)+\frac{4}{3}T(R)N_{\mathrm{f}}\right]
   \frac{1}{\epsilon}+O(g^4)
   \right\}.
\label{eq:(A1)}
\end{equation}


The fermion mass:
\begin{equation}
   m_0=m
   \left[
   1+\frac{g^2}{(4\pi)^2}C_2(R)(-3)
   \frac{1}{\epsilon}+O(g^4)
   \right].
\label{eq:(A2)}
\end{equation}

The gauge potential (in the Feynman gauge):
\begin{equation}
   A_\mu^a(x)=\left[1+\frac{g^2}{(4\pi)^2}C_2(G)(-1)
   \frac{1}{\epsilon}+O(g^4)\right]A_{\mu R}^a(x).
\label{eq:(A3)}
\end{equation}

The fermion field:
\begin{equation}
   \psi(x)=
   \left[1+\frac{g^2}{(4\pi)^2}C_2(R)\left(-\frac{1}{2}\right)
   \frac{1}{\epsilon}+O(g^4)\right]
   \psi_R(x).
\label{eq:(A4)}
\end{equation}

\subsection{Composite operators}
The bare operators~\eqref{eq:(4.7)}--\eqref{eq:(4.11)} and renormalized
counterparts
\begin{align}
   \left\{\mathcal{O}_{1\mu\nu}\right\}_R(x)&\equiv
   \left\{F_{\mu\rho}^aF_{\nu\rho}^a\right\}_R(x),
\label{eq:(A5)}\\
   \left\{\mathcal{O}_{2\mu\nu}\right\}_R(x)&\equiv
   \delta_{\mu\nu}\left\{F_{\rho\sigma}^aF_{\rho\sigma}^a\right\}_R(x),
\label{eq:(A6)}\\
   \left\{\mathcal{O}_{3\mu\nu}\right\}_R(x)&\equiv
   \left\{\Bar{\psi}
   \left(\gamma_\mu\overleftrightarrow{D}_\nu
   +\gamma_\nu\overleftrightarrow{D}_\mu\right)
   \psi\right\}_R(x),
\label{eq:(A7)}\\
   \left\{\mathcal{O}_{4\mu\nu}\right\}_R(x)&\equiv
   \delta_{\mu\nu}\left\{\Bar{\psi}
   \overleftrightarrow{\Slash{D}}
   \psi\right\}_R(x),
\label{eq:(A8)}\\
   \left\{\mathcal{O}_{5\mu\nu}\right\}_R(x)&\equiv
   \delta_{\mu\nu}m\left\{\Bar{\psi}
   \psi\right\}_R(x)
\label{eq:(A9)}
\end{align}
are related as
\begin{equation}
   \mathcal{O}_{i\mu\nu}(x)
   =Z_{ij}\left\{\mathcal{O}_{j\mu\nu}\right\}_R(x).
\label{eq:(A10)}
\end{equation}
The gluonic contribution to the operator renormalization
of~$\mathcal{O}_{1\mu\nu}$ and~$\mathcal{O}_{2\mu\nu}$ was determined
in~Ref.~\cite{Suzuki:2013gza}. By further computing fermionic contributions to
the operator renormalization (corresponding to diagrams~A18 and~A19 in the main
text) and taking the gauge coupling and wave function normalizations
[Eqs.~\eqref{eq:(A1)} and~\eqref{eq:(A3)}] into account, we have
\begin{align}
   Z_{11}
   &=1+\frac{g^2}{(4\pi)^2}C_2(G)
   \left(-\frac{11}{3}\right)
   \frac{1}{\epsilon}+O(g^4),
\label{eq:(A11)}\\
   Z_{12}
   &=\frac{g^2}{(4\pi)^2}C_2(G)\frac{11}{12}\frac{1}{\epsilon}+O(g^4),
\label{eq:(A12)}\\
   Z_{13}
   &=\frac{g^4}{(4\pi)^2}C_2(R)\frac{2}{3}\frac{1}{\epsilon}+O(g^6),
\label{eq:(A13)}\\
   Z_{14}
   &=\frac{g^4}{(4\pi)^2}C_2(R)\left(-\frac{1}{3}\right)\frac{1}{\epsilon}
   +O(g^6),
\label{eq:(A14)}\\
   Z_{15}
   &=\frac{g^4}{(4\pi)^2}C_2(R)(-3)\frac{1}{\epsilon}+O(g^6),
\label{eq:(A15)}
\end{align}
and
\begin{align}
   Z_{21}
   &=0,
\label{eq:(A16)}\\
   Z_{22}
   &=1+O(g^4),
\label{eq:(A17)}\\
   Z_{23}
   &=0,
\label{eq:(A18)}\\
   Z_{24}
   &=O(g^6),
\label{eq:(A19)}\\
   Z_{25}
   &=\frac{g^4}{(4\pi)^2}C_2(R)(-12)\frac{1}{\epsilon}+O(g^6).
\label{eq:(A20)}
\end{align}
From these, to the one-loop order, we further have
\begin{align}
   &\delta_{\mu\nu}\left\{F_{\mu\rho}^aF_{\nu\rho}^a\right\}_R(x)
\notag\\
   &=\left[
   1+\frac{g^2}{(4\pi)^2}C_2(G)\left(\frac{11}{6}\right)\right]
   \left\{F_{\rho\sigma}^aF_{\rho\sigma}^a\right\}_R(x)
\notag\\
   &\qquad{}
   +\frac{g^4}{(4\pi)^2}C_2(R)\left(-\frac{2}{3}\right)
   \left\{\Bar{\psi}\overleftrightarrow{\Slash{D}}\psi\right\}_R(x)
   +\frac{g^4}{(4\pi)^2}C_2(R)(-6)
   m\left\{\Bar{\psi}\psi\right\}_R(x).
\label{eq:(A21)}
\end{align}

On the other hand, the computation of diagrams B03, B04, B05, B06, and~B07,
combined with the wave function renormalization~\eqref{eq:(A4)}, shows
\begin{align}
   Z_{31}
   &=\frac{1}{(4\pi)^2}T(R)N_{\mathrm{f}}
   \frac{16}{3}
   \frac{1}{\epsilon}+O(g^2),
\label{eq:(A22)}\\
   Z_{32}
   &=\frac{1}{(4\pi)^2}T(R)N_{\mathrm{f}}
   \left(-\frac{4}{3}\right)
   \frac{1}{\epsilon}+O(g^2),
\label{eq:(A23)}\\
   Z_{33}
   &=1+\frac{g^2}{(4\pi)^2}C_2(R)\left(-\frac{8}{3}\right)\frac{1}{\epsilon}
   +O(g^4),
\label{eq:(A24)}\\
   Z_{34}
   &=\frac{g^2}{(4\pi)^2}C_2(R)\frac{4}{3}\frac{1}{\epsilon}
   +O(g^4),
\label{eq:(A25)}\\
   Z_{35}
   &=O(g^4),
\label{eq:(A26)}
\end{align}
and
\begin{equation}
   Z_{41}=Z_{43}=0,\qquad Z_{42}=Z_{45}=O(g^4),\qquad Z_{44}=1+O(g^4).
\label{eq:(A27)}
\end{equation}
The consistency of these relations shows
\begin{align}
   &\delta_{\mu\nu}\left\{\Bar{\psi}\gamma_\mu\overleftrightarrow{D}_\nu
   \psi\right\}_R(x)
\notag\\
   &=\left[1+\frac{g^2}{(4\pi)^2}C_2(R)\frac{4}{3}\right]
   \left\{\Bar{\psi}\overleftrightarrow{\Slash{D}}
   \psi\right\}_R(x)
   +\frac{1}{(4\pi)^2}T(R)N_{\mathrm{f}}\left(-\frac{4}{3}\right)
   \left\{F_{\rho\sigma}^aF_{\rho\sigma}^a\right\}_R(x).
\label{eq:(A28)}
\end{align}

Finally, a general theorem (see, for example, Ref.~\cite{Collins:1984xc}) says
that
\begin{equation}
   Z_{51}=Z_{52}=Z_{53}=Z_{54}=0,\qquad Z_{55}=1.
\label{eq:(A29)}
\end{equation}

\section{Integration formulas}
\label{sec:B}
\begin{equation}
   \int_\ell\frac{1}{(\ell^2)^\alpha}\mathrm{e}^{-s\ell^2}
   =\frac{\Gamma(D/2-\alpha)}{(4\pi)^{D/2}\Gamma(D/2)}s^{\alpha-D/2}.
\label{eq:(B1)}
\end{equation}

\begin{equation}
   \int_k\int_\ell\frac{\mathrm{e}^{-sk^2-u\ell^2-v(k+\ell)^2}}{k^2}
   =\frac{1}{(4\pi)^D(D/2-1)(u+v)}(su+uv+vs)^{1-D/2}.
\label{eq:(B2)}
\end{equation}

\begin{align}
   &\int_\ell \mathrm{e}^{-s\ell^2}=\frac{1}{(4\pi)^{D/2}}s^{-D/2}.
\label{eq:(B3)}\\
   &\int_\ell \mathrm{e}^{-s\ell^2}\ell_\mu\ell_\nu
   =\frac{1}{(4\pi)^{D/2}}s^{-D/2-1}\frac{1}{2}\delta_{\mu\nu}.
\label{eq:(B4)}\\
   &\int_\ell \mathrm{e}^{-s\ell^2}
   \ell_\mu\ell_\nu\ell_\rho\ell_\sigma
   =\frac{1}{(4\pi)^{D/2}}s^{-D/2-2}
   \frac{1}{4}\left(
   \delta_{\mu\nu}\delta_{\rho\sigma}
   +\delta_{\mu\rho}\delta_{\nu\sigma}
   +\delta_{\mu\sigma}\delta_{\nu\rho}
   \right).
\label{eq:(B5)}\\
   &\int_\ell \mathrm{e}^{-s\ell^2}
   \ell_\mu\ell_\nu\ell_\rho\ell_\sigma\ell_\alpha\ell_\beta
   =\frac{1}{(4\pi)^{D/2}}s^{-D/2-3}\frac{1}{8}
   \left(
   \delta_{\mu\nu}\delta_{\rho\sigma}\delta_{\alpha\beta}
   +\text{$14$ permutations}\right).
\label{eq:(B6)}
\end{align}

\begin{align}
   &\int_\ell\frac{1}{\ell^2}\,
   \mathrm{e}^{-s\ell^2}=\frac{1}{(4\pi)^{D/2}}\frac{1}{D/2-1}s^{-D/2+1}.
\label{eq:(B7)}\\
   &\int_\ell\frac{1}{\ell^2}\,\mathrm{e}^{-s\ell^2}\ell_\mu\ell_\nu
   =\frac{1}{(4\pi)^{D/2}}\frac{1}{D}s^{-D/2}\delta_{\mu\nu}.
\label{eq:(B8)}\\
   &\int_\ell\frac{1}{\ell^2}\,\mathrm{e}^{-s\ell^2}
   \ell_\mu\ell_\nu\ell_\rho\ell_\sigma
   =\frac{1}{(4\pi)^{D/2}}\frac{1}{2(D+2)}s^{-D/2-1}
   \left(
   \delta_{\mu\nu}\delta_{\rho\sigma}
   +\delta_{\mu\rho}\delta_{\nu\sigma}
   +\delta_{\mu\sigma}\delta_{\nu\rho}
   \right).
\label{eq:(B9)}\\
   &\int_\ell\frac{1}{\ell^2}\,\mathrm{e}^{-s\ell^2}
   \ell_\mu\ell_\nu\ell_\rho\ell_\sigma\ell_\alpha\ell_\beta
\notag\\
   &=\frac{1}{(4\pi)^{D/2}}\frac{1}{4(D+4)}s^{-D/2-2}
   \left(
   \delta_{\mu\nu}\delta_{\rho\sigma}\delta_{\alpha\beta}
   +\text{$14$ permutations}\right).
\label{eq:(B10)}
\end{align}

\section{Gluonic contributions to~$\zeta_{1j}^{(1)}(t)$}
\label{sec:C}
Gluonic contributions to~$\zeta_{1j}^{(1)}(t)$ are tabulated
in~Table~\ref{table:0}.
\begin{table}
\caption{$\zeta_{1j}^{(1)}$ in units of $\frac{g_0^2}{(4\pi)^2}C_2(G)$.
The numbers with~$*$ are corrected from previous versions of the present paper.
}
\label{table:0}
\begin{center}
\renewcommand{\arraystretch}{2.2}
\setlength{\tabcolsep}{20pt}
\begin{tabular}{crr}
\toprule
 diagram & \multicolumn{1}{c}{$\zeta_{11}^{(1)}(t)$}
 & \multicolumn{1}{c}{$\zeta_{12}^{(1)}(t)$} \\
\midrule
A03  & $0$ & $0$ \\
A04  & $-3\epsilon(t)^{-1}-1$ & $0$ \\
A05  & $-\dfrac{7}{36}$ & $-\dfrac{49}{144}$ \\
A06  & $2\epsilon(t)^{-1}-\dfrac{1}{2}$ & $0$ \\
A07  & $\dfrac{19}{288}$ & $\dfrac{121}{384}$ \\
A08  & $\dfrac{35}{96}^*$ & $\dfrac{143}{384}^*$ \\
A09  & $-\dfrac{25}{8}$ & $0$ \\
A11  & $\dfrac{1}{3}\epsilon(t)^{-1}-\dfrac{17}{36}$ & $\dfrac{7}{12}\epsilon(t)^{-1}+\dfrac{1}{144}$ \\
A13  & $-\dfrac{5}{3}\epsilon(t)^{-1}+\dfrac{25}{36}$ & $-\dfrac{3}{2}\epsilon(t)^{-1}-\dfrac{29}{16}$ \\
A14  & $3\epsilon(t)^{-1}+3$ & $0$ \\
A15  & $3\epsilon(t)^{-1}+\dfrac{5}{2}$ & $0$ \\
A16  & $1^*$ & $\dfrac{31}{24}$ \\
\bottomrule
\end{tabular}
\end{center}
\end{table}

\section{Justification of our computational prescription}
\label{sec:D}
In this appendix, we give detailed explanation why the one-loop matching
coefficients~$\zeta^{(1)}_{ij}(t)$ in~Eq.~\eqref{eq:(4.25)} can be determined
without computing correlation functions~Eqs.~\eqref{eq:(4.26)}
and~\eqref{eq:(4.29)} with operators~$\Tilde{\mathcal{O}}_{i\mu\nu}(t,x)$ are
replaced by corresponding bare operators
in~Eqs.~\eqref{eq:(4.6)}--\eqref{eq:(4.10)}, that is,
\begin{equation}
   \left\langle\mathcal{O}_{i\mu\nu}(x)
   A_\beta^b(y)A_\gamma^c(z)\right\rangle,
\label{eq:(D1)}
\end{equation}
and
\begin{equation}
   \left\langle\mathcal{O}_{i\mu\nu}(x)
   \psi(y)\Bar{\psi}(z)\right\rangle.
\label{eq:(D2)}
\end{equation}
We argue that one can neglect contributions of these correlation functions
altogether, if one follows a regularization prescription for IR divergences we
adopted in the main text.

Both Eqs.~\eqref{eq:(D1)} and~\eqref{eq:(D2)} can be treated in a similar
manner, so we consider Eq.~\eqref{eq:(D1)}. Let us take a particular one-loop
1PI Feynman diagram that contributes to~Eq.~\eqref{eq:(D1)}, for instance,
diagram~A11 in~Fig.~\ref{fig:22}. This is a diagram in the $D$-dimensional
gauge theory and thus consists only of ordinary (i.e., filled circle) vertices
and ordinary (i.e., no Gaussian factor in~Eq.~\eqref{eq:(3.17)}) propagators.
Writing the contribution of this diagram to~Eq.~\eqref{eq:(D1)}
as~Eq.~\eqref{eq:(4.27)}, the vertex
part~$\mathcal{M}_{\mu\nu,\beta\gamma}(k,\ell)$ is given by a one-loop integral,
\begin{equation}
   \mathcal{M}_{\mu\nu,\beta\gamma}(k,\ell)=\int_p\,
   \mathcal{I}_{\mu\nu,\beta\gamma}(p;k,\ell),
\label{eq:(D3)}
\end{equation}
where dimensional counting says that the
integrand~$\mathcal{I}_{\mu\nu,\beta\gamma}(p;k,\ell)$ is of mass dimension~$-2$.
In general, this integral is UV divergent and, if we further Taylor-expand the
integrand with respect to the external momenta $k$ and~$\ell$, exhibits also IR
divergences.

Next, we note that for the above 1PI diagram in the $D$-dimensional gauge
theory, there always exists a corresponding flow Feynman diagram with the same
structure (i.e., diagram~A11 in~Fig.~\ref{fig:22} in the present example). We
write its contribution to the vertex
part~$\mathcal{M}_{\mu\nu,\beta\gamma}(k,l)$ in~Eq.~\eqref{eq:(4.27)} as
\begin{equation}
   \mathcal{M}_{\mu\nu,\beta\gamma}(k,\ell)=\int_p\,
   \mathcal{I}_{\mu\nu,\beta\gamma}(p;k,\ell;t),
\label{eq:(D4)}
\end{equation}
where the Feynman rules for the vertices are completely identical to those
for~Eq.~\eqref{eq:(D3)}, while propagators are identical except the
Gaussian factor in~Eq.~\eqref{eq:(3.18)} which depends on the flow time of
the inserted composite operator~$\Tilde{\mathcal{O}}_{i\mu\nu}(t,x)$; we have
explicitly indicated the dependence of the integrand on~$t$.

Now, in the \emph{one-loop level}, it can be seen that what is relevant
to~$\zeta^{(1)}_{ij}(t)$ is the difference between Eqs.~\eqref{eq:(D4)}
and~\eqref{eq:(D3)}:
\begin{equation}
   \int_p\,\left[
   \mathcal{I}_{\mu\nu,\beta\gamma}(p;k,\ell;t)
   -\mathcal{I}_{\mu\nu,\beta\gamma}(p;k,\ell)\right].
\label{eq:(D5)}
\end{equation}
From the structure of two integrands explained above, (logarithmic) IR
divergences are cancelled out in the difference and we may Taylor-expand the
integrand with respect to the external momenta~$k$ and~$\ell$. The coefficient
of the $O(k,\ell)$-term is then given by
\begin{equation}
   C\int_p\frac{\mathrm{e}^{-2tp^2}-1}{(p^2)^2}
   =C\frac{1}{(4\pi)^{D/2}}\frac{4}{(D-2)(D-4)}(2t)^{2-D/2},
\label{eq:(D6)}
\end{equation}
where $C$ is a constant and a complex dimension~$D$ is introduced \emph{to
regularize UV divergences}; the last expression has been obtained by the
analytic continuation from the complex domain~$2<\re(D)<4$. The
combination~\eqref{eq:(D6)} (with $D\to4$) is relevant for the contribution of
the diagram under consideration to~$\zeta_{ij}^{(1)}$.

Now, although the above computation is a proper one, there exists a ``facile
method'' that reproduces Eq.~\eqref{eq:(D6)} with much less effort; the
argument proceeds as follows.

One considers only the contribution of the flow Feynman diagram~\eqref{eq:(D4)}
and Taylor-expand the integrand with respect to the external momenta~$k$
and~$\ell$. In this way, one encounters IR divergences. One then introduces a
complex dimension~$D$ \emph{to regularize IR divergences}. Note that this is
possible because the integral~\eqref{eq:(D4)} is UV finite thanks to the
Gaussian factors; there always exists a complex domain of~$D$ with which the
integral is well-defined. Then the coefficient of the $O(k,\ell)$-term is given
by
\begin{equation}
   C\int_p\frac{\mathrm{e}^{-2tp^2}}{(p^2)^2}
   =C\frac{1}{(4\pi)^{D/2}}\frac{4}{(D-2)(D-4)}(2t)^{2-D/2}.
\label{eq:(D7)}
\end{equation}
This is an expression obtained by the analytic continuation from the complex
domain~$4<\re(D)$. The last expression is however identical
to~Eq.~\eqref{eq:(D6)}.

Thus, we have arrived at the following facile method: We completely forget
about the computation of correlation functions with bare operators,
Eqs.~\eqref{eq:(D1)} and~\eqref{eq:(D2)}. For correlation functions with flowed
operators, Eqs.~\eqref{eq:(4.26)} and~\eqref{eq:(4.29)}, we simply
Taylor-expand the integrand of the Feynman integral with respect to the
external momenta (and the fermion mass). Resulting IR divergences are
regularized by ``dimensional regularization'', i.e., the analytic continuation
from~$\re(D)>4$. Then the result for the one-loop matching
coefficients~$\zeta_{ij}^{(1)}$ is identical to the one obtained by a
computation that properly takes the contribution of Eqs.~\eqref{eq:(D1)}
and~\eqref{eq:(D2)} into account. This facile method is precisely the
computational method we made use of in the main text.

\section{Modified energy--momentum tensor}
\label{sec:E}
The argument in~Sec.~\ref{sec:2} shows that the energy--momentum
tensor~\eqref{eq:(2.7)} fulfills the Ward--Takahashi relations,
\begin{equation}
   \left\langle\partial_\mu T_{\mu\nu}(x)A_\rho(y)\dotsm\right\rangle
   =-\delta(x-y)
   \left\langle\left[\partial_\nu A_\rho(y)
   -D_\rho A_\nu(y)\right]\dotsm\right\rangle+\dotsb,
\label{eq:(E1)}
\end{equation}
and
\begin{equation}
   \left\langle\partial_\mu T_{\mu\nu}(x)\psi(y)\dotsm\right\rangle
   =-\delta(x-y)
   \left\langle D_\nu \psi(y)\dotsm\right\rangle+\dotsb.
\label{eq:(E2)}
\end{equation}
In particular, there is no term being proportional
to~$\partial_\nu\delta(x-y)$ in the right-hand sides. These relations lead to
\begin{align}
   &\left\langle\partial_\mu\left[x_\nu T_{\mu\nu}(x)\right]
   A_\rho(y)\dotsm\right\rangle
\notag\\
   &=\left\langle T_{\mu\mu}(x)
   A_\rho(y)\dotsm\right\rangle
   -x_\nu\delta(x-y)
   \left\langle\left[\partial_\nu A_\rho(y)
   -D_\rho A_\nu(y)\right]\dotsm\right\rangle+\dotsb,
\label{eq:(E3)}
\end{align}
and
\begin{equation}
   \left\langle\partial_\mu\left[x_\nu T_{\mu\nu}(x)\right]
   \psi(y)\dotsm\right\rangle
   =\left\langle T_{\mu\mu}(x)
   \psi(y)\dotsm\right\rangle
   -x_\nu\delta(x-y)
   \left\langle D_\nu \psi(y)\dotsm\right\rangle+\dotsb.
\label{eq:(E4)}
\end{equation}
Integrating these over the whole space, we have
\begin{align}
   \int\mathrm{d}^Dx\,
   \left\langle T_{\mu\mu}(x)
   A_\rho(y)\dotsm\right\rangle
   &=\left\langle\left[y_\nu\partial_\nu A_\rho(y)
   -y_\nu D_\rho A_\nu(y)\right]\dotsm\right\rangle+\dotsb
\notag\\
   &=\left\langle\left\{\left(y_\nu\partial_\nu+1\right) A_\rho(y)
   -D_\rho\left[y_\nu A_\nu(y)\right]\right\}\dotsm\right\rangle+\dotsb,
\label{eq:(E5)}
\end{align}
and
\begin{align}
   \int\mathrm{d}^Dx\,\left\langle T_{\mu\mu}(x)
   \psi(y)\dotsm\right\rangle
   &=\left\langle y_\nu D_\nu \psi(y)\dotsm\right\rangle+\dotsb
\notag\\
   &=\left\langle\left[y_\nu\partial_\nu\psi(y)+y_\nu A_\nu(y)\psi(y)\right]
   \dotsm\right\rangle+\dotsb.
\label{eq:(E6)}
\end{align}

Now, we see that Eq.~\eqref{eq:(E5)} is accord with the naively-expected
relation,
\begin{equation}
   \int\mathrm{d}^Dx\,\left\langle T_{\mu\mu}(x)
   \mathcal{O}(y)\dotsm\right\rangle
   =\left\langle\left(y_\nu\partial_\nu+d_\mathcal{O}\right)
   \mathcal{O}(y)\dotsm\right\rangle+\dotsb,
\label{eq:(E7)}
\end{equation}
where $d_{\mathcal{O}}$ is the scale dimension of the field~$\mathcal{O}$, up to
the gauge transformation, but Eq.~\eqref{eq:(E6)} is not. One would expect
instead
\begin{equation}
   \int\mathrm{d}^Dx\,\left\langle\Tilde{T}_{\mu\mu}(x)
   \psi(y)\dotsm\right\rangle
   =\left\langle\left(y_\nu\partial_\nu+\frac{D-1}{2}\right)\psi(y)
   \dotsm\right\rangle+\dotsb,
\label{eq:(E8)}
\end{equation}
up to the gauge transformation. One can construct such a modified
energy--momentum tensor by using a freedom to add the equation of motion to
the energy--momentum tensor. We see that Eq.~\eqref{eq:(E8)} is realized, if
\begin{equation}
   \left\langle\partial_\mu\Tilde{T}_{\mu\nu}(x)\psi(y)\dotsm\right\rangle
   =-\delta(x-y)
   \left\langle D_\nu \psi(y)\dotsm\right\rangle
   +\frac{D-1}{2D}\partial_\nu^x\delta(x-y)
   \left\langle\psi(y)\dotsm\right\rangle+\dotsb.
\label{eq:(E9)}
\end{equation}
It can be seen that the last term does not influence on Ward--Takahashi
relations associated with the translation and the rotation (Poincar\'e
transformations). This modification is accomplished by
\begin{equation}
   \Tilde{T}_{\mu\nu}(x)=T_{\mu\nu}(x)
   +\frac{D-1}{2D}
   \delta_{\mu\nu}
   \Bar{\psi}(x)\left(\overleftrightarrow{\Slash{D}}+2m_0\right)\psi(x).
\label{eq:(E10)}
\end{equation}
This modified energy--momentum tensor, fulfilling the naively-expected
relation~\eqref{eq:(E7)}, leads to a simpler derivation of the trace anomaly.
In terms of flowed operators, this modification amounts to adding the
combination~\eqref{eq:(5.1)} times
\begin{equation}
   \frac{D-1}{2D}\to\frac{3}{8}.
\end{equation}
We note however that when the energy--momentum tensor is separated from other
operators in position space, the modification has no effect because it is
proportional to the equation of motion.


\begin{thebibliography}{00}

\bibitem{Callan:1970ze} 
  C.~G.~Callan, Jr., S.~R.~Coleman, and R.~Jackiw,
  Ann.\ Phys.\  {\bf 59}, 42 (1970).

\bibitem{Coleman:1970je} 
  S.~R.~Coleman and R.~Jackiw,
  Ann.\ Phys.\  {\bf 67}, 552 (1971).

\bibitem{Freedman:1974gs} 
  D.~Z.~Freedman, I.~J.~Muzinich and E.~J.~Weinberg,
  Ann.\ Phys.\  {\bf 87}, 95 (1974).

\bibitem{Joglekar:1975jm} 
  S.~D.~Joglekar,
  Ann.\ Phys.\  {\bf 100}, 395 (1976);
  {\bf 102}, 594 (1976) [erratum].

\bibitem{Caracciolo:1988hc} 
  S.~Caracciolo, G.~Curci, P.~Menotti and A.~Pelissetto,
  Nucl.\ Phys.\ B {\bf 309}, 612 (1988).

\bibitem{Caracciolo:1989pt} 
  S.~Caracciolo, G.~Curci, P.~Menotti and A.~Pelissetto,
  Ann.\ Phys.\  {\bf 197}, 119 (1990).

\bibitem{Suzuki:2013gza} 
  H.~Suzuki,
  PTEP {\bf 2013}, 083B03 (2013)
  [arXiv:1304.0533 [hep-lat]].

\bibitem{Luscher:2010iy} 
  M.~L\"uscher,
  JHEP {\bf 1008}, 071 (2010)
  [arXiv:1006.4518 [hep-lat]].

\bibitem{Luscher:2011bx} 
  M.~L\"uscher and P.~Weisz,
  JHEP {\bf 1102}, 051 (2011)
  [arXiv:1101.0963 [hep-th]].

\bibitem{Luscher:2013cpa} 
  M.~L\"uscher,
  JHEP {\bf 1304}, 123 (2013)
  [arXiv:1302.5246 [hep-lat]].

\bibitem{Giusti:2010bb} 
  L.~Giusti and H.~B.~Meyer,
  Phys.\ Rev.\ Lett.\  {\bf 106}, 131601 (2011)
  [arXiv:1011.2727 [hep-lat]].

\bibitem{Giusti:2012yj} 
  L.~Giusti and H.~B.~Meyer,
  JHEP {\bf 1301}, 140 (2013)
  [arXiv:1211.6669 [hep-lat]].

\bibitem{Robaina:2013zmb} 
  D.~Robaina and H.~B.~Meyer,
  PoS LATTICE {\bf 2013}, 323 (2014)
  [arXiv:1310.6075 [hep-lat]].

\bibitem{Giusti:2013sqa} 
  L.~Giusti and H.~B.~Meyer,
  PoS LATTICE {\bf 2013}, 214 (2013)
  [arXiv:1310.7818 [hep-lat]].

\bibitem{Giusti:2013mxa} 
  L.~Giusti and M.~Pepe,
  arXiv:1311.1012 [hep-lat].

\bibitem{Giusti:2014ila} 
  L.~Giusti and M.~Pepe,
  Phys.\ Rev.\ Lett.\  {\bf 113}, 031601 (2014)
  [arXiv:1403.0360 [hep-lat]].

\bibitem{Suzuki:2012wx} 
  H.~Suzuki,
  Phys.\ Lett.\ B {\bf 719}, 435 (2013)
  [arXiv:1209.5155 [hep-lat]].

\bibitem{Luscher:2013vga} 
  M.~L\"uscher,
  PoS LATTICE {\bf 2013}, 016 (2014)
  [arXiv:1308.5598 [hep-lat]].

\bibitem{Borsanyi:2012zs} 
  S.~Bors\'anyi, S.~D\"urr, Z.~Fodor, C.~Hoelbling, S.~D.~Katz, S.~Krieg, T.~Kurth and L.~Lellouch {\it et al.},
  JHEP {\bf 1209}, 010 (2012)
  [arXiv:1203.4469 [hep-lat]].

\bibitem{Borsanyi:2012zr} 
  S.~Bors\'anyi, S.~D\"urr, Z.~Fodor, S.~D.~Katz, S.~Krieg, T.~Kurth, S.~Mages and A.~Sch\"afer {\it et al.},
  arXiv:1205.0781 [hep-lat].

\bibitem{Fodor:2012td} 
  Z.~Fodor, K.~Holland, J.~Kuti, D.~Nogradi, and C.~H.~Wong,
  JHEP {\bf 1211}, 007 (2012)
  [arXiv:1208.1051 [hep-lat]].

\bibitem{Fritzsch:2013je} 
  P.~Fritzsch and A.~Ramos,
  JHEP {\bf 1310}, 008 (2013)
  [arXiv:1301.4388 [hep-lat]].

\bibitem{Monahan:2013lwa} 
  C.~Monahan and K.~Orginos,
  PoS Lattice {\bf 2013}, 443 (2014)
  [arXiv:1311.2310 [hep-lat]].

\bibitem{Shindler:2013bia} 
  A.~Shindler,
  Nucl.\ Phys.\ B {\bf 881}, 71 (2014)
  [arXiv:1312.4908 [hep-lat]].

\bibitem{Bonati:2014tqa} 
  C.~Bonati and M.~D'Elia,
  Phys.\ Rev.\ D {\bf 89}, 105005 (2014)
  [arXiv:1401.2441 [hep-lat]].

\bibitem{DelDebbio:2013zaa} 
  L.~Del Debbio, A.~Patella, and A.~Rago,
  JHEP {\bf 1311}, 212 (2013)
  [arXiv:1306.1173 [hep-th]].

\bibitem{Asakawa:2013laa} 
  M.~Asakawa {\it et al.}  [FlowQCD Collaboration],
  Phys.\ Rev.\ D {\bf 90}, no. 1, 011501 (2014)
  [arXiv:1312.7492 [hep-lat]].

\bibitem{Collins:1984xc} 
  J.~C.~Collins,
  {\it Renormalization.\ An Introduction to Renormalization, the Renormalization Group, and the Operator Product Expansion\/}
  (Cambridge University Press, Cambridge, 1984).

\bibitem{Crewther:1972kn} 
  R.~J.~Crewther,
  Phys.\ Rev.\ Lett.\  {\bf 28}, 1421 (1972).

\bibitem{Chanowitz:1972vd} 
  M.~S.~Chanowitz and J.~R.~Ellis,
  Phys.\ Lett.\ B {\bf 40}, 397 (1972).

\bibitem{Adler:1976zt} 
  S.~L.~Adler, J.~C.~Collins and A.~Duncan,
  Phys.\ Rev.\ D {\bf 15}, 1712 (1977).

\bibitem{Nielsen:1977sy} 
  N.~K.~Nielsen,
  Nucl.\ Phys.\ B {\bf 120}, 212 (1977).

\bibitem{Collins:1976yq} 
  J.~C.~Collins, A.~Duncan, and S.~D.~Joglekar,
  Phys.\ Rev.\ D {\bf 16}, 438 (1977).

\bibitem{Fujikawa:1980rc} 
  K.~Fujikawa,
  Phys.\ Rev.\ D {\bf 23}, 2262 (1981).

\bibitem{Fujikawa:2004cx} 
  K.~Fujikawa and H.~Suzuki,
  {\it Path Integrals and Quantum Anomalies\/} (Clarendon, Oxford, 2004).

\bibitem{Caswell:1974gg} 
  W.~E.~Caswell,
  Phys.\ Rev.\ Lett.\  {\bf 33}, 244 (1974).

\bibitem{Jones:1974mm} 
  D.~R.~T.~Jones,
  Nucl.\ Phys.\ B {\bf 75}, 531 (1974).

\bibitem{Tarrach:1980up} 
  R.~Tarrach,
  Nucl.\ Phys.\ B {\bf 183}, 384 (1981).

\bibitem{Nachtmann:1981zg} 
  O.~Nachtmann and W.~Wetzel,
  Nucl.\ Phys.\ B {\bf 187}, 333 (1981).

\bibitem{Nakamura:2004sy} 
  A.~Nakamura and S.~Sakai,
  Phys.\ Rev.\ Lett.\  {\bf 94}, 072305 (2005)
  [hep-lat/0406009].

\bibitem{Meyer:2007ic} 
  H.~B.~Meyer,
  Phys.\ Rev.\ D {\bf 76}, 101701 (2007)
  [arXiv:0704.1801 [hep-lat]].

\bibitem{Meyer:2007dy} 
  H.~B.~Meyer,
  Phys.\ Rev.\ Lett.\  {\bf 100}, 162001 (2008)
  [arXiv:0710.3717 [hep-lat]].

\bibitem{Capitani:1998mq} 
  S.~Capitani {\it et al.}  [ALPHA Collaboration],
  Nucl.\ Phys.\ B {\bf 544}, 669 (1999)
  [hep-lat/9810063].

\bibitem{Itou:2013faa} 
  E.~Itou,
  PoS LATTICE {\bf 2013}, 005 (2014)
  [arXiv:1311.2676 [hep-lat]].

\bibitem{Fodor:2014pqa} 
  Z.~Fodor, K.~Holland, J.~Kuti, D.~Nogradi and C.~H.~Wong,
  PoS LATTICE {\bf 2013}, 062 (2014)
  [arXiv:1401.2176 [hep-lat]].

\end{thebibliography}
\end{document}